\documentclass[reprint,amsmath,amssymb]{revtex4-1}
\usepackage[utf8x]{inputenc}
\usepackage{graphicx}
\usepackage{dcolumn}
\usepackage{bm}
\usepackage{color}

\usepackage{soul} 
\usepackage[normalem]{ulem} 

\begin{document}

\preprint{APS/123-QED}

\title{Geometric cumulants associated with adiabatic cycles crossing
  degeneracy points: Application to finite size scaling of
  metal-insulator transitions in crystalline electronic systems}

\author{Bal\'azs Het\'enyi$^{1,2,3}$ and Sertaç Cengiz$^3$}

\affiliation{ $^1$Department of Theoretical Physics, Institute of
  Physics, Budapest University of Technology and Economics, H-1111
  Budapest, Hungary \\
  $^2$MTA-BME Topology and Correlations Research
  Group, Department of Theoretical Physics, Budapest University of
  Technology and Economics, H-1521 Budapest, Hungary \\
  $^3$Department of Physics, Bilkent University, TR-06800 Bilkent, Ankara,
  Turkey}

\date{\today}

\begin{abstract}
In this work we focus on two questions.  One, we complement the
machinary to calculate geometric phases along adiabatic cycles as
follows.  The geometric phase is a line integral along an adiabatic
cycle, and if the cycle encircles a degeneracy point, the phase
becomes non-trivial.  If the cycle crosses the degeneracy point the
phase diverges.  We construct quantities which are well-defined when
the path crosses the degeneracy point.  We do this by constructing a
generalized Bargmann invariant, and noting that it can be interpreted
as a cumulant generating function, with the geometric phase being the
first cumulant.  We show that particular ratios of cumulants remain
finite for cycles crossing a set of isolated degeneracy points.  The
cumulant ratios take the form of the Binder cumulants known from the
theory of finite size scaling in statistical mechanics (we name them
geometric Binder cumulants).  Two, we show that the machinery
developed can be applied to perform finite size scaling in the context
of the modern theory of polarization.  The geometric Binder cumulants
are size independent at gap closure points or regions with closed gap
(Luttinger liquid).  We demonstrate this by model calculations for a
one-dimensional topological model, several two-dimensional models, and
a one-dimensional correlated model.  In the case of two dimensions we
analyze to different situations, one in which the Fermi surface is
one-dimensional (a line), and two cases in which it is zero
dimensional (Dirac points).  For the geometric Binder cumulants the
gap closure points can be found by one dimensional scaling even in two
dimensions.  As a technical point we stress that only certain finite
difference approximations for the cumulants are applicable, since not
all approximation schemes are capable of extracting the size scaling
information in the case of a closed gap system.
\end{abstract}

\pacs{}

\maketitle

\section{Introduction}

Berry's geometric phase~\cite{Berry84} is an integral of a connection
along a quantum adiabatic cycle.  Nontrivial values arise when the
curve corresponding to the adiabatic cycle encircles a degeneracy
point.  The geometric phase can be viewed~\cite{Souza00} as the first
member of a series of cumulants extracted from the evolution of the
wave function along the adiabatic cycle.  The higher order cumulants
are not geometric, they depend on the parametrization of the cycle,
not only on its geometry, but they are also physically well-defined.
We show that particular ratios of the cumulants give finite values for
cycles crossing degeneracy points.  When the path crosses a degeneracy
point, the Berry phase becomes undefined, the higher order cumulants
all diverge.  The machinery can not be applied in this case.  In this
work we complement the formalism of the geometric phase by
constructing quantities which are meaningful in the case of paths
which cross degeneracy (or gap closure) points.

Loosely speaking, the physical situation can be compared to
electrodynamics in materials~\cite{Imada98}.  An insulator is a gapped
system, and can be described by quantities such as the polarization,
linear and nonlinear dielectric susceptibilities.  In a gapless
system, the polarization is undefined and the dielectric
susceptibility diverges with system size.  The relevant quantities
become the current, the Drude weight, or the conductivity.  As we will
show, a similar dichotomy exists for adiabatic paths: a gapped path
can be described by the Berry phase, and the gauge invariant
cumulants.  When gap closure occurs along the path, the relevant
quantities become a particular set of cumulant ratios.

A variant of the Berry phase, known as the Zak phase~\cite{Zak89},
obtained by integrating across the Brillouin zone of a crystalline
system, plays the central role in the modern theory of
polarization~\cite{King-Smith93,Resta94,Resta00,Vanderbilt18} (MTP).
The Zak phase is proportional to the bulk polarization of a
crystalline system.  It is of interest to note that our understanding
of conduction and insulation in quantum systems was initiated by the
seminal work of Kohn~\cite{Kohn64}, who emphasized that the classical
idea of localization of single charge carriers as a criterion of
insulation loses its validity.  Instead, the criterion of insulation,
according to Kohn, is many-body localization, the localization of the
center of mass of the entire charge distribution (or at least the
charge distribution of large chunks of the sample).  In systems with
open boundaries, Kudinov~\cite{Kudinov91} showed that the variance of
the polarization (which is proportional to the variance of the center
of mass of the charge distribution) is an appropriate criterion to
distinguish conductors from insulators.  For crystalline systems,
however, it was not possible to calculate the center of mass of the
charge distribution, nor its variance or higher cumulants, because the
relevant quantum mechanical operator, the position operator, is
ill-defined under periodic boundary conditions.  It was precisely this
problem that was
solved~\cite{King-Smith93,Resta94,Resta00,Vanderbilt18} by MTP, by
casting the polarization in crystallie systems as a Zak phase.

As an extension of the MTP, Souza, Wilkens, and Martin~\cite{Souza00}
introduced the so called gauge invariant cumulants, which are
essential in the study of
polarization~\cite{Marzari97,Marzari12,Watanabe18,Yahyavi17} as well
as charge transport, since they provide access to important related
quantities, such as the variance of the polarization~\cite{Resta99},
or the shift current~\cite{Patankar18} (related to the third
cumulant).  Patankar et al.~\cite{Patankar18} use the ratio of the
third cumulant and the second as a gauge of nonlinearity in the
Su-Schrieffer-Heeger model~\cite{Su79}.  The Zak phase~\cite{Zak89} is
also the starting point to construct topological invariants, the
quantities which characterize the different phases of topological
insulators~\cite{Bernevig13,Haldane88,Kane13,Kane05a,Kane05b}.  The
MTP formalism has also been applied~\cite{Thonhauser06} to study the
topological Haldane model.

A limitation of the quantities derived from the MTP is that at gap
closure information about system size scaling is lost.  The variance
of the polarization, as derived by Resta, diverges at gap closure,
even for finite systems.  For this reason, techniques based on the
finite size scaling hypothesis~\cite{Fisher72a,Fisher72b} are not
directly applicable to metal-insulator transitions.  Application of
the Binder cumulant technique~\cite{Binder81a,Binder81b} (based on the
finite size scaling hypothesis) in MTP is difficult for two reasons:
other than the loss of size scaling information already mentioned, a
Binder cumulant is a ratio of averages of expectation values of
observables (the order parameter), and in MTP the basic quantity is a
geometric phase, rather than the expectation value of an observable.
In this work we overcome both of these obstacles.

Cumulants are logarithmic derivatives of characteristic functions
(also known as cumulant generating functions).  Since there exist
cumulants in the context of the Zak phase, we first show that for
general geometric phases (not necessarily a Zak phase) a
generalization of the Bargmann invariant \cite{Bargmann64} plays the
role of the characteristic function.  The generalized Bargmann
invariant is most closely related to the characteristic function of a
periodic probability distribution, since moments and cumulants can be
obtained from it via {\it finite difference derivatives}.

Our construction of cumulant ratios is similar to that of the Binder
cumulants~\cite{Binder81a,Binder81b}, for this reason, we will use the
term geometric Binder cumulant (GBC).  The GBCs are zero if the
adiabatic path encounters no gap closure points, otherwise they are
finite.  As a preliminary example, we calculate the GBC for a
spin-$\frac{1}{2}$ particle in a precessing magnetic field.  We then
implement our construction in the
MTP~\cite{King-Smith93,Resta94,Resta00,Vanderbilt18,Thonhauser06} and
show that the GBCs are an effective tool for finite size scaling of
metal-insulator~\cite{Imada98} and other quantum phase transitions.
Gap closure points (or regions) can be located.  As a technicality, we
emphasize the use of a particular application of finite difference
derivatives which guarantee the correct scaling with system size in
gapless systems.

We also investigate the applicability of the MTP to two-dimensional
and one-dimensional interacting systems.  In the former, the set of
gap closure points can be one or zero dimensional (Dirac points in
graphene).  The question we pose is whether the cumulants constructed
within the MTP can signal a zero dimensional gap closure.  We find in
cases such as graphene (or the Haldane model) scaling has to be done
in one dimension only.  As for the interacting system we study, the
Berry phase corresponding to the polarization is a single-point Berry
phase.  Nevertheless, the gap closure region is located via
application of the GBC.

In section \ref{sec:stat}, for background, we introduce the
characteristic function and show how moments and cumulants are
generated from it.  We emphasize the distinction between a general and
a periodic probability distribution function.  In the former, moments
and cumulants are obtained via derivatives, while in the latter one
has to resort to finite difference derivatives.  In section
\ref{sec:geoBinder} we use the Bargmann invariant~\cite{Bargmann64} as
a starting point to derive the GBCs.  In section \ref{sec:spinhalf}
basic calculations are presented for a spin-$\frac{1}{2}$ particle in
a rotating magnetic field.  In section \ref{sec:pol} we construct GBCs
for the finite scaling of the polarization, emphasizing the importance
of an alternative approximation scheme which gives the correct size
scaling information at gap closure.  In section
\ref{sec:ideal_conductor} we give expressions for the variance of the
polarization and the GBC at different approximation levels for an
ideal conductor (gap closure point).  In section
\ref{sec:Model_Calculations} we present model calculations for one and
two-dimensional band insulators, as well as for a correlated system.
In section \ref{sec:cnclsn} we conclude our work.

\section{Background: moments, central moments, cumulants and Binder cumulants}
\label{sec:stat}

To initiate the discussion we give an overview of some quantities used
in statistics to characterize probability distributions.  Given a real
random variable $x$ and a probability distribution $P(x)$ which
satisfies
\begin{equation}
  P(x) \geq 0; \int_{-\infty}^{\infty}d x P(x) = 1.
\end{equation}
The characteristic function $f(k)$ is the Fourier transform of $P(x)$,
\begin{equation}
  \label{eqn:f_k}
  f(k) = \int_{-\infty}^{\infty}d x P(x) e^{i k x}.
\end{equation}
The $n$th moment of the distribution $P(x)$ is defined as
\begin{equation}
  \label{eqn:M_n}
  M_n = \frac{1}{i^n} \left. \frac{\partial^n f(k)}{\partial k^n} \right|_{k=0} = \langle x^n \rangle.
\end{equation}
The $n$th cumulant of $P(x)$ is defined as
\begin{equation}
  \label{eqn:C_n}
  C_n = \frac{1}{i^n} \left. \frac{\partial^n \ln f(k)}{\partial k^n} \right|_{k=0}.
\end{equation}
Cumulants can be written in terms of moments (and vice versa).  For
the first four cumulants the expressions in terms of moments are:
\begin{eqnarray}
C_1 &=& M_1,  \\ \nonumber
C_2 &=& M_2 - M_1^2, \\ \nonumber
C_3 &=& M_3 - 3M_2 M_1 + 2 M_1^3, \\ \nonumber
C_4 &=& M_4 - 4M_3 M_1 - 3 M_2^2 + 12 M_2M_1^2 - 6 M_1^4.
\end{eqnarray}
The first four cumulants are named as follows: $C_1$ the mean, $C_2$
the variance, $C_3$ the skew, and $C_4$ is the kurtosis.  Cumulants of
order higher than one are independent of the mean or the origin of the
coordinate system.

It is possible to define the central moments, which are the moments
derived from the probability distribution $P(x) \rightarrow P(x +
M_1)$, shifted to give a zero first moment.  $M_1$ is the first moment
of the unshifted distribution.  Using the shifted probability
distribution in Eq. (\ref{eqn:C_n}) to derive cumulants results in
\begin{eqnarray}
C_1 &=& 0,  \\ \nonumber
C_2 &=& M_2, \\ \nonumber
C_3 &=& M_3, \\ \nonumber
C_4 &=& M_4 - 3 M_2^2.  
\end{eqnarray}
Note that if $P(x)$ is shifted, it is the phase of the corresponding
characteristic function that shifts,
\begin{equation}
  f(k) \rightarrow f(k)e^{-i k M_1}.
\end{equation}

We now consider a distribution periodic in $L$,
\begin{equation}
  \label{eqn:px_L}
  P_L(x) = \sum_{w = -\infty}^{\infty} P(x + w L).
\end{equation}
In this case, the characteristic function associated with $P_L(x)$ is
discrete,
\begin{equation}
  \label{eqn:f_q}
  f_q = \int_0^L d x P_L(x) e^{i 2 \pi q x / L}; q \in \mathbb{Z}.
\end{equation}
Substituting Eq. (\ref{eqn:px_L}) into Eq. (\ref{eqn:f_q}) we see that
\begin{equation}
  f_q = f\left( \frac{2 \pi}{L} q \right),
\end{equation}
where the right hand side is the characteristic function of the
distribution $P(x)$ (Eq. (\ref{eqn:f_k})) evaluated at $k = \frac{2
  \pi}{L} q$.

Since for periodic distributions the characteristic function is only
well-defined at a discrete set of $k$-points, the moments and
cumulants have to be based on finite difference derivatives of $f_q$,
rather than continuous derivatives.  We define them as follows:
\begin{equation}
  \label{eqn:M_n_f}
  M_n = \begin{cases}
    (-1)^{\frac{n-1}{2}}\left(\frac{L}{2\pi}\right)^n  \mbox{Im} \left. \frac{\mathbb{D}^n f_q}{\mathbb{D} q^n} \right|_{q=0} & \mbox{if $n$ is odd}, \\
    \hspace{.3cm} (-1)^{\frac{n}{2}}\left(\frac{L}{2\pi}\right)^n  \mbox{Re} \left. \frac{\mathbb{D}^n f_q}{\mathbb{D} q^n} \right|_{q=0} & \mbox{if $n$ is even}.
    \end{cases}
\end{equation}
\begin{equation}
  \label{eqn:C_n_f}
  C_n = \begin{cases}
    (-1)^{\frac{n-1}{2}}\left(\frac{L}{2\pi}\right)^n  \mbox{Im} \left. \frac{\mathbb{D}^n \ln f_q}{\mathbb{D} q^n} \right|_{q=0} & \mbox{if $n$ is odd}, \\
    \hspace{.3cm} (-1)^{\frac{n}{2}}\left(\frac{L}{2\pi}\right)^n  \mbox{Re} \left. \frac{\mathbb{D}^n \ln f_q}{\mathbb{D} q^n} \right|_{q=0} & \mbox{if $n$ is even}.
    \end{cases}
\end{equation}
where the notation $\frac{\mathbb{D}}{\mathbb{D} q}$ denotes a finite
difference derivative.  There are different types of finite difference
approximations~\cite{Hildebrand68}.  We will adhere to the convention
introduced by Resta and Sorella in the context of the modern theory of
polarization, and use central difference approximations in all cases.
Unless otherwise indicated, we use the lowest order approximation, but
in sections \ref{sec:ideal_conductor} and \ref{sec:Model_Calculations}
we will make comparisons between approximations of different orders.

Applying the lowest order finite difference approximation results in
\begin{eqnarray}
  \label{eqn:C_12_L}
C_1 &=& \frac{L}{2\pi} \mbox{Im} \frac{\ln f_1 - \ln f_{-1}}{2} = \frac{L}{2\pi} \mbox{Im} \ln f_1,\\ \nonumber
C_2 &=& -\left(\frac{L}{2\pi}\right)^2 \mbox{Re} (\ln f_1 + \ln f_{-1} - 2 \ln f_0) \\ &=&
-\frac{L^2}{2\pi^2} \mbox{Re} \ln f_1, \nonumber
\end{eqnarray}
where we used that $f_q = f_{-q}^*$ and that $f_0 = 1$.  Note the
similarity of the $C_1$ and $C_2$ in Eq. (\ref{eqn:C_12_L}) to the
Resta~\cite{Resta98} and Resta-Sorella~\cite{Resta99} expressions for
the polarization and its variance.  In the limit $L\rightarrow \infty$
$C_1$ and $C_2$ converge to $\langle x \rangle$ and $\langle x^2
\rangle - \langle x \rangle^2$, respectively, and higher order
cumulants constructed via Eq. (\ref{eqn:C_n}) also reproduce the
cumulants (Eq. (\ref{eqn:C_n})).

The Binder cumulant~\cite{Binder81a,Binder81b} is a quantity used to
locate phase transition points and critical exponents.  It is an
application of the finite size scaling
hypothesis~\cite{Fisher72a,Fisher72b}.  In numerical simulations the
thermodynamic limit is not accessible, for this reason, phase
transition points can be shifted or smeared out when numerically
calculated based on susceptibilities or other response functions.
These numerical artifacts are eliminated at phase transition points by
the use of the Binder cumulant.  The method is useful in
classical~\cite{Selke06} as well as quantum~\cite{Hetenyi99} phase
transitions, however, heretofore the construction has only been
applied in phase transitions characterized by order parameters,
i.e. expectation values of Hermitian operators.

In the construction due to Binder, particular ratios of moments (or
cumulants) are taken.  The Binder cumulants take known values at phase
transition points which are independent of system size.  It is this
property which makes the Binder cumulant a computationally useful tool
in locating critical points.  One commonly used Binder cumulant is
\begin{equation}
  \label{eqn:U4}
  U_4 = 1 - \frac{1}{3} \frac{M_4}{M_2^2} = -\frac{1}{3}\frac{C_4}{C_2^2}.
\end{equation}
The crucial point is that the product of the orders of moments in the
numerator and the denominator are equal.  This is the reason the size
dependence cancels at critical points.

\section{Moments, central moments, cumulants, and Binder cumulants associated with adiabatic cycles}

\label{sec:geoBinder}

In this section we derive the Berry phase.  We show that the Berry
phase can be viewed as the first in a cumulant sequence associated
with the Bargmann invariant.  An extended version of the Bargmann
invariant will take the place of the characteristic function, and from
it the above mentioned cumulant sequence can be derived through the
application of finite difference derivatives, as was done in
Eqs. (\ref{eqn:M_n_f}) and (\ref{eqn:C_n_f}).  We also show that
ratios of moments constructed á la Binder are physically well defined
geometric quantities (the GBC).  In subsequent sections we show that
they take non-trivial values when the adiabatic path encounters a
degeneracy point.

By adiabatic cycle we mean a cycle for which the adiabatic theorem
holds.  This theorem was first proven by Born and Fock~\cite{Born28}
for the nondegenerate case.  They considered paths along which the
particular state under consideration is gapped, as well as paths along
which level crossings occur at a finite number of singular points
along the path.  Kato~\cite{Kato50} generalized the results of Born
and Fock to the case of degenerate states.  The paths considered in
this case are such that a set of degenerate states are isolated from
the rest of the Hilbert space of the system via an energy gap, but,
again, a finite number of gap closures with other sets of degenerate
states can occur at crossing points.

To derive the Berry phase and cumulants valid for the general case, we
consider a parameter space $\vec{\bf \xi}$.  At each point of this
parameter space a Hamiltonian is defined, for which it is valid that
\begin{equation}
  H(\vec{\bf \xi}) |\Psi_n(\vec{\bf \xi})\rangle = E_n(\vec{\bf \xi}) |\Psi_n(\vec{\bf \xi})\rangle,
\end{equation}
where $E_n(\vec{\bf \xi})$($|\Psi_n(\vec{\bf \xi})\rangle$) denote
energy eigenvalues(eigenstates).  We will consider the ground state,
$|\Psi_0(\vec{\bf \xi})\rangle$, but the construction is general.  We
consider a set of points in the parameter space, $\vec{\bf \xi}_m$
with $m = 1,...,N$, and form the discrete Berry phase based on the
cyclic product,
\begin{equation}
  \label{eqn:Z_qa}
  \tilde{Z}_q = \prod_{m=1}^N \langle \Psi(\vec{\bf \xi}_m) | \Psi(\vec{\bf \xi}_{m+q})\rangle,
\end{equation}
where $\vec{\bf \xi}_{N+s}$ is set equal to $\vec{\bf \xi}_s$, and
where the index for the ground state, subscript $0$, was suppressed.
The cyclic product $\tilde{Z}_1$ is known as the Bargmann
invariant~\cite{Bargmann64}.  $\tilde{Z}_q$ is the general discrete
cumulant generating function, an extension of the Bargmann invariant.
The series of moments or cumulants can be generated from $\tilde{Z}_q$
by setting $f_q = \tilde{Z}_q$, and using Eqs.  (\ref{eqn:M_n_f}) and
(\ref{eqn:C_n_f}) and by applying finite difference derivative
formulas with respect to the discrete variable $q$.  The interval in
this case is not $frac{L}{2\pi}$, but unity.

The first cumulant that results from this procedure is the discrete
Berry phase,
\begin{equation}
  \label{eqn:C1}
  C_1 = \mbox{Im} \left. \frac{\mathbb{D} \ln Z_q}{\mathbb{D} q} \right|_{q=0} = \mbox{Im} (\ln \tilde{Z}_1 - \ln \tilde{Z}_0) = \mbox{Im} \ln \tilde{Z}_1,
\end{equation}
since $\tilde{Z}_0 = 1$.  Some of the higher order cumulants may be
written,
\begin{eqnarray}
  \label{eqn:cmlnts_adiabatic}
  C_2 &=& - 2 \mbox{Re} \ln \tilde{Z}_1 \\ \nonumber
  C_3 &=&-\mbox{Im} \ln \tilde{Z}_2 + 2 \mbox{Im} \ln \tilde{Z}_1, \\ \nonumber
  C_4 &=& 2 \mbox{Re} \ln \tilde{Z}_2 - 8 \mbox{Re} \ln \tilde{Z}_1.
\end{eqnarray}
Odd cumulants are sums of phases of $\tilde{Z}_q$, even ones are sums
of the logarithms of their magnitudes.  The
continuous Berry phase can be derived from Eqs. (\ref{eqn:Z_qa}) and
(\ref{eqn:C1}) by assuming that the parameter series $\vec{\bf \xi}_m$
is placed along a closed curve in the order $m=1,...,N$.  We can take
the continuous limit by increasing $N$, the number of points
representing the closed curve (simultaneously decrease the distances
between them).  We can then use the expansion up to first order
\begin{equation}
  | \Psi(\vec{\bf \xi}_{m+1})\rangle = | \Psi(\vec{\bf \xi}_m)\rangle + \delta {\vec{\xi}} \cdot \nabla_{\vec{\bf \xi}} | \Psi(\vec{\bf \xi}_ m)\rangle,
\end{equation}
to rewrite $C_1$ as
\begin{equation}
  \label{eqn:C1_xi}
  C_1 = \mbox{Im} \sum_{m=1}^N \delta {\vec{\xi}} \cdot \langle
  \Psi(\vec{\bf \xi}_m)| \nabla_{\vec{\bf \xi}} | \Psi(\vec{\bf
    \xi}_m)\rangle,
\end{equation}
or, in the continuous limit,
\begin{equation}
  \label{eqn:C1_xi_cont}
  C_1 = \mbox{Im} \oint d \vec{\bf \xi} \cdot \langle \Psi(\vec{\bf \xi}) |\nabla_{\vec{\bf \xi}}| \Psi(\vec{\bf \xi}) \rangle.
\end{equation}

It is useful to introduce a parametrization for the closed curve in
parameter space as $\vec{\xi}(t)$, where $t\in [0,T]$.  We now
discretize as $t_m = m \delta t$ (with $\delta t = \frac{T}{N}$).  In
this case, the discrete path Berry phase (\ref{eqn:C1_xi}) can be
written as
\begin{equation}
  \label{eqn:C1_t}
  C_1 = \mbox{Im} \sum_{m=1}^N \delta t \langle \Psi(t_m) |
  \partial_t | \Psi(t_m)\rangle,
\end{equation}
or in the continuous limit,
\begin{equation}
  \label{eqn:C1_t_cont}
  C_1 = \mbox{Im} \int_0^T d t \langle \Psi(t) | \partial_t | \Psi(t)\rangle.
\end{equation}
Expanding $\ln \tilde{Z}_1$ to second order, and using the definition of $C_2$
(Eq. (\ref{eqn:cmlnts_adiabatic})) results in,
\begin{eqnarray}
  \label{eqn:C2_xi}
  C_2 &=& - \sum_{m=1}^N \left[ \langle \Psi (\vec{\bf \xi}_m)|(\delta {\vec{\xi}} \cdot \nabla_{\vec{\bf \xi}})^2|\Psi (\vec{\bf \xi}_m) \rangle \right. \\ & &-
\left.  \langle \Psi (\vec{\bf \xi}_m)|(\delta {\vec{\xi}} \cdot \nabla_{\vec{\bf \xi}})|\Psi (\vec{\bf \xi}_m) \rangle^2 \right], \nonumber
\end{eqnarray}
or using the parametrization
\begin{eqnarray}
  \label{eqn:C2_t}
  C_2 &=& - \delta t^2 \sum_{m=1}^N \left[ \langle \Psi (t_m)|\partial_t^2|\Psi (t_m) \rangle \right. \\ & &-
\left.  \langle \Psi (t_m)|\partial_t|\Psi (t_m) \rangle^2\right]. \nonumber
\end{eqnarray}
The continuous limit can only be taken in Eq. (\ref{eqn:C2_t}), not in
Eq. (\ref{eqn:C2_xi}).  In Eq. (\ref{eqn:C2_t}) one factor of $\delta
t$ has to be eliminated, and only then do we obtain
\begin{eqnarray}
  \label{eqn:C2_t_cont}
  \Gamma_2 &=& - \int_0^T d t \left[ \langle \Psi (t)|\partial_t^2|\Psi (t) \rangle \right. \\ & &-
\left.  \langle \Psi (t)|\partial_t|\Psi (t) \rangle^2\right]. \nonumber
\end{eqnarray}
$\Gamma_2$ is physically well-defined, however, it is no longer
geometric.  It is not merely a function of the cyclic path, it also
depends on ``how fast'' the path is traversed in the variable $t$.
$\Gamma_2$ gives finite numbers for adiabatic paths which do not
encounter crossing points.  The situation is similar for all higher
order cumulants.  An $n$th order $C_n$ will lead to a factor of
$\delta t^n$, from which $n-1$ have to be eliminated to lead to a
$\Gamma_n$ which is physically well-defined, albeit, not geometric.
Cumulants constructed this way are known as gauge invariant cumulants,
introduced in the context of the MTP by Souza, Wilkens, and
Martin~\cite{Souza00}.

\begin{table*}[t]
\begin{center}
\begin{tabular}{ |c|c|c|c| } 
  \hline Characteristic & Definition & Associated & Moments, cumulants  \\ function & & with &  \\
  \hline \hline
  & & & \\
  $f(k)$ & $= \int_{-\infty}^{\infty} d x \exp(i k x)P(x)$ & $P(x)$, probability distribution & $M_n = \frac{1}{i^n} \left. \frac{\partial^n f(k)}{\partial k^n} \right|_{k=0}$\\
  & & of a continuous random variable & $C_n = \frac{1}{i^n} \left. \frac{\partial^n \ln f(k)}{\partial k^n} \right|_{k=0}$\\
  & & & \\ \hline
  & & & \\
  $f_q$ & $=\int_0^L d x \exp\left(i \frac{2 \pi x}{L} q \right) P_L(x)$ & $P_L(x)$, probability distribution & $M_n = \begin{cases}
    (-1)^{\frac{n-1}{2}}\left(\frac{L}{2\pi}\right)^n  \mbox{Im} \left. \frac{\mathbb{D}^n f_q}{\mathbb{D} q^n} \right|_{q=0} & \mbox{if $n$ is odd}, \\
    \hspace{.3cm} (-1)^{\frac{n}{2}}\left(\frac{L}{2\pi}\right)^n  \mbox{Re} \left. \frac{\mathbb{D}^n f_q}{\mathbb{D} q^n} \right|_{q=0} & \mbox{if $n$ is even}.
    \end{cases}$\\
  & $q$ integer & of a continuous random variable &   $C_n = \begin{cases}
    (-1)^{\frac{n-1}{2}}\left(\frac{L}{2\pi}\right)^n  \mbox{Im} \left. \frac{\mathbb{D}^n \ln f_q}{\mathbb{D} q^n} \right|_{q=0} & \mbox{if $n$ is odd}, \\
    \hspace{.3cm} (-1)^{\frac{n}{2}}\left(\frac{L}{2\pi}\right)^n  \mbox{Re} \left. \frac{\mathbb{D}^n \ln f_q}{\mathbb{D} q^n} \right|_{q=0} & \mbox{if $n$ is even}.
    \end{cases}$
\\
  & & periodic in $L$ & \\
  & & & \\ \hline
  & & & \\
  $\tilde{Z}_q$ & $=\prod_{s=1}^N \langle \Psi(\xi_s)| \Psi(\xi_{s+q}\rangle$ &
  geometric phases & $M_n = \begin{cases}
    (-1)^{\frac{n-1}{2}} \mbox{Im} \left. \frac{\mathbb{D}^n \tilde{Z}_q}{\mathbb{D} q^n} \right|_{q=0} & \mbox{if $n$ is odd}, \\
    \hspace{.3cm} (-1)^{\frac{n}{2}} \mbox{Re} \left. \frac{\mathbb{D}^n \tilde{Z}_q}{\mathbb{D} q^n} \right|_{q=0} & \mbox{if $n$ is even}.
    \end{cases}$ \\
  & $q$ integer & along adiabatic cycles & $C_n = \begin{cases}
    (-1)^{\frac{n-1}{2}} \mbox{Im} \left. \frac{\mathbb{D}^n \ln \tilde{Z}_q}{\mathbb{D} q^n} \right|_{q=0} & \mbox{if $n$ is odd}, \\
    \hspace{.3cm} (-1)^{\frac{n}{2}}  \mbox{Re} \left. \frac{\mathbb{D}^n \ln \tilde{Z}_q}{\mathbb{D} q^n} \right|_{q=0} & \mbox{if $n$ is even}.
    \end{cases}$ \\
  & & & \\ \hline
  & & & \\
  $Z_q$ & $=\langle \Psi_0| \exp \left( i \frac{2 \pi \hat{X}}{L}q
  \right)| \Psi_0\rangle$ & polarization distribution & $M_n = \begin{cases}
    (-1)^{\frac{n-1}{2}}\left(\frac{L}{2\pi}\right)^n  \mbox{Im} \left. \frac{\mathbb{D}^n Z_q}{\mathbb{D} q^n} \right|_{q=0} & \mbox{if $n$ is odd}, \\
    \hspace{.3cm} (-1)^{\frac{n}{2}}\left(\frac{L}{2\pi}\right)^n  \mbox{Re} \left. \frac{\mathbb{D}^n Z_q}{\mathbb{D} q^n} \right|_{q=0} & \mbox{if $n$ is even}.
    \end{cases}$\\
  & $q$ integer & of many-body systems & $C_n = \begin{cases}
    (-1)^{\frac{n-1}{2}}\left(\frac{L}{2\pi}\right)^n  \mbox{Im} \left. \frac{\mathbb{D}^n \ln Z_q}{\mathbb{D} q^n} \right|_{q=0} & \mbox{if $n$ is odd}, \\
    \hspace{.3cm} (-1)^{\frac{n}{2}}\left(\frac{L}{2\pi}\right)^n  \mbox{Re} \left. \frac{\mathbb{D}^n \ln Z_q}{\mathbb{D} q^n} \right|_{q=0} & \mbox{if $n$ is even}.
    \end{cases}$\\
  & & periodic in $L$ & \\
  & &  &\\ \hline
  & &  & \\
  $Z_q^{(b)}$ & $= \mbox{Det}\prod_{s=1}^L S(k_s,k_{s+q})$ & polarization
  distribution & $M_n = \begin{cases}
    (-1)^{\frac{n-1}{2}}\left(\frac{L}{2\pi}\right)^n  \mbox{Im} \left. \frac{\mathbb{D}^n Z_q^{(b)}}{\mathbb{D} q^n} \right|_{q=0} & \mbox{if $n$ is odd}, \\
    \hspace{.3cm} (-1)^{\frac{n}{2}}\left(\frac{L}{2\pi}\right)^n  \mbox{Re} \left. \frac{\mathbb{D}^n Z_q^{(b)}}{\mathbb{D} q^n} \right|_{q=0} & \mbox{if $n$ is even}.
    \end{cases}$\\
  & $S(k_s,k_{s+q}) = \langle u_{k_s,m} | u_{k_{s+q},m'} \rangle$ & of band systems & \\
  & $q$ integer, & periodic in $L$ & $C_n = \begin{cases}
    (-1)^{\frac{n-1}{2}}\left(\frac{L}{2\pi}\right)^n  \mbox{Im} \left. \frac{\mathbb{D}^n \ln Z_q^{(b)}}{\mathbb{D} q^n} \right|_{q=0} & \mbox{if $n$ is odd}, \\
    \hspace{.3cm} (-1)^{\frac{n}{2}}\left(\frac{L}{2\pi}\right)^n  \mbox{Re} \left. \frac{\mathbb{D}^n \ln Z_q^{(b)}}{\mathbb{D} q^n} \right|_{q=0} & \mbox{if $n$ is even}.
    \end{cases}$ \\ 
  & $m,m'$ occupied bands & & \\ 
  & $| u_{k_s,m'} \rangle$ periodic Bloch state &  & \\
  & &  & \\ \hline
   \hline
\end{tabular}
\label{tab:chf}
\caption{Different types of characteristic functions (cumulant
  generating functions) discussed in this work.  The definitions of
  the characteristic functions are given in the second column.  The
  third column specifies the contexts in which they appear.  The
  fourth column specifies how moments ($M_n$) and cumulants ($C_n$)
  are obtained from the characteristic functions.  The notation
  $\frac{\mathbb{D}^n ...}{\mathbb{D} q^n}$ denotes an $n$th finite
  difference derivative.}
\end{center}
\end{table*}

It is possible to construct quantities which are gauge invariant as
well as geometric.  Let us first write $C_4$ as
\begin{eqnarray}
  \label{eqn:C4_t}
  C_4 &=& \delta t^4 \sum_{m=1}^N \left[ \langle \Psi (t_m)|\partial_t^4|\Psi (t_m) \rangle \right. \\
    \nonumber
    & & \left. -4 \langle \Psi (t_m)|\partial_t^3|\Psi (t_m) \rangle\langle \Psi (t_m)|\partial_t|\Psi (t_m) \rangle \right. \\
    \nonumber
    & & \left. -3 \langle \Psi (t_m)|\partial_t^2|\Psi (t_m) \rangle\langle \Psi (t_m)|\partial_t^2|\Psi (t_m) \rangle \right. \\
    \nonumber
    & & \left. +12 \langle \Psi (t_m)|\partial_t^2|\Psi (t_m) \rangle\langle \Psi (t_m)|\partial_t|\Psi (t_m) \rangle^2    \right. \\
    \nonumber
    & & \left. -6 \langle \Psi (t_m)|\partial_t|\Psi (t_m) \rangle^4\right]. 
\end{eqnarray}
The quantity
\begin{equation}
  \label{eqn:U4new}
  U_4 = -\lim_{\delta t \rightarrow 0}\frac{1}{3}\frac{C_4}{C_2^2},
\end{equation}
is gauge invariant and geometric, where $C_4$ and $C_2$ are defined as
in Eqs. (\ref{eqn:C4_t}) and (\ref{eqn:C2_t}), respectively.  In
Eq. (\ref{eqn:U4new}) the $\delta t$s cancel, because, as is the case
in the original Binder cumulants, the products of the orders of the
cumulants in the numerator and the denominator are equal.

If the adiabatic path does not encounter a gap closure point, then
$U_4 = 0$.  Defining $C_4 = \delta t^3 \Gamma_4$, and taking the limit
$\delta t \rightarrow 0$ results in
\begin{equation}
  \label{eqn:U4Gamma}
  U_4 = \lim_{\delta t \rightarrow 0}\delta t\frac{\Gamma_4}{\Gamma_2^2} \rightarrow 0.
\end{equation}
($\Gamma_4$ does take finite values for adiabatic paths which do not
encounter degeneracy points.)  However, $U_4$ gives a finite value if
the adiabatic path {\it does} encounter a degeneracy point.

\section{Spin-$\frac{1}{2}$ particle in a rotating magnetic field}
\label{sec:spinhalf}

In this section we calculate $U_4$ for a spin-$\frac{1}{2}$ particle
in a rotating magnetic field.  This system was used as an example in
the original work of Berry~\cite{Berry84} to demonstrate the meaning
of the Berry phase arising through adiabatic paths which encircle the
degeneracy.  Here we show what happens when the path touches the
degeneracy point.

The Hamiltonian of a spin-$\frac{1}{2}$ particle in a magnetic field
is given by
\begin{equation}
  H = - g \frac{e}{2 m_e}\vec{\bf S} \cdot \vec{\bf B},
\end{equation}
where 
\begin{equation}
  \vec{\bf S} = \frac{\hbar}{2} \vec{\bf \sigma},
\end{equation}
and $\vec{\bf \sigma}$ is a vector of Pauli matrices, $\sigma_x$,
$\sigma_y$, and $\sigma_z$.  We assume that the magnetic field
$\vec{\bf B}$ is time-dependent and precesses at frequency $\omega$
according to
\begin{eqnarray}
  B_x & = & B \sin(\theta) \cos (\omega t) \\ \nonumber
  B_y & = & B \sin(\theta) \sin (\omega t) \\ \nonumber
  B_z & = & B \cos(\theta).
\end{eqnarray}
We set $\phi = \omega t$.  This system can be solved exactly.  The
solution we use to construct moments has the form
\begin{equation}
  |\Psi_- (\theta,\phi)\rangle = \left( { \begin{array}{c} -e^{i \phi} \sin\left(\frac{\theta}{2}\right) \\ \cos\left(\frac{\theta}{2}\right) \end{array} } \right).
\end{equation}
Using $|\Psi_- (\theta,\phi)\rangle$ we construct a $Z_q$ along a path
in the space $\vec{\bf B}$ discretized according to $\phi_m = m
\frac{2 \pi}{N}$, with $m=1,...,N$, resulting in,
\begin{equation}
  \label{eqn:Zqspin}
  \tilde{Z}_q = \left( \sin^2 \frac{\theta}{2} e^{i \frac{2 \pi}{N}q} + \cos^2 \frac{\theta}{2} \right)^N.
\end{equation}
From Eq. (\ref{eqn:M_n_f}) it can be shown that
\begin{eqnarray}
  M_4 &=& 2 (|\tilde{Z}_2| - 4 |\tilde{Z}_1| + 3)  \\ \nonumber
  M_2 &=& 2 (|\tilde{Z}_1| - 1 ), 
\end{eqnarray}
(the magnitudes of all the $\tilde{Z}_q$s were taken, to center the
``distribution'') and using Eq. (\ref{eqn:Zqspin}) it follows that
\begin{equation}
  \frac{M_4}{M_2^2} = 3,
\end{equation}
implying that $U_4$, as defined in Eq. (\ref{eqn:U4}) is zero.

The gap closure of this system occurs at the origin.  We can also
consider a cyclic path which crosses this point.  In this case, since
for all $n \neq 0$ $Z_n=0$, and $Z_0 = 1$, we obtain,
\begin{equation}
  \frac{M_4}{M_2^2} = \frac{3}{2},
\end{equation}
implying that $U_4$ is $\frac{1}{2}$.  The GBC gives a finite number
for a cyclic path which {\it encounters} the degeneracy point.

The above result is valid for any parametrization of the adiabatic
loop.  The fact that the GBC is finite for a loop which encounters the
degeneracy point is due to the fact that in the product on the
right-hand side of Eq. (\ref{eqn:Z_qa}) the scalar product ``across''
the degeneracy point gives zero, and therefore the entire product will
be zero.  Reparametrizing the loop will not change this.  Also, the
direction of the path as it crosses the degeneracy point is not
important either, since the result that $n \neq 0$ $Z_n=0$, and $Z_0 =
1$ is independent of direction.

\section{Cumulants of the polarization}
\label{sec:pol}

In this section, we make the connection between the cumulant machinery
and the MTP explicit.  We first show that the quantity which plays the
central role in MTP, sometimes called~\cite{Watanabe18} the
polarization amplitude (defined below in Eq. (\ref{eqn:Z_q})), is a
discrete characteristic function of the type derived in
Eq. (\ref{eqn:f_q}).  As such, the cumulants~\cite{Resta98,Resta99}
can only be derived via finite difference derivatives.  Second, we
show that if we evaluate the polarization amplitude for a band
insulator (where the wave function is a Slater determinant), then it
takes the form of the generalized Bargmann invariant defined in
Eq. (\ref{eqn:Z_qa}).  This derivation links the quantities introduced
in Sections \ref{sec:stat} and \ref{sec:geoBinder}.

We consider a many-electron system, one-dimensional for convenience,
periodic in $L$.  In our calculations below, $L$ will always refer to
the size of the supercell of the system, in which periodic boundary
conditions are imposed.  In band systems $2\pi/L$ is the spacing
between k-points in the Brillouin zone.  We write the quantity known
as the polarization amplitude as,
\begin{equation}
  \label{eqn:Z_q}
  Z_q = \langle \Psi_0 |\exp \left( i \frac{2 \pi q}{L} \hat{X}
  \right) |\Psi_0 \rangle,
\end{equation}
where $\Psi_0$ denotes the ground state wave function of the system,
$q$ is an integer, and the total position operator is defined as
\begin{equation}
  \hat{X} = \sum_{j=1}^L \hat{n}_j j,
\end{equation}
where $\hat{n}_j$ is the density operator at site $j$.

To show that $Z_q$ is the analog of a characteristic function we can
write the distribution of the total position corresponding to the
state $\Psi_0$ 
\begin{equation}
  P_L(X) = \langle \Psi_0 |\delta(\hat{X}-X)| \Psi_0\rangle.
\end{equation}
The distribution $P_L(X)$ is manifestly periodic in $L$.  We can write
$Z_q$ as
\begin{equation}
  Z_q = \int d X P_L(X) \exp \left( i \frac{2 \pi q}{L} X \right).
\end{equation}
It is obvious that $Z_q$ is a discrete characteristic function of the
type given in Eq. (\ref{eqn:f_q}).  We can derive the $n$th moment or
cumulant of $X$ by applying Eqs. (\ref{eqn:M_n_f}) and
(\ref{eqn:C_n_f}).
\begin{figure}[ht]
  \centering
 \includegraphics[width=6cm,keepaspectratio=true]{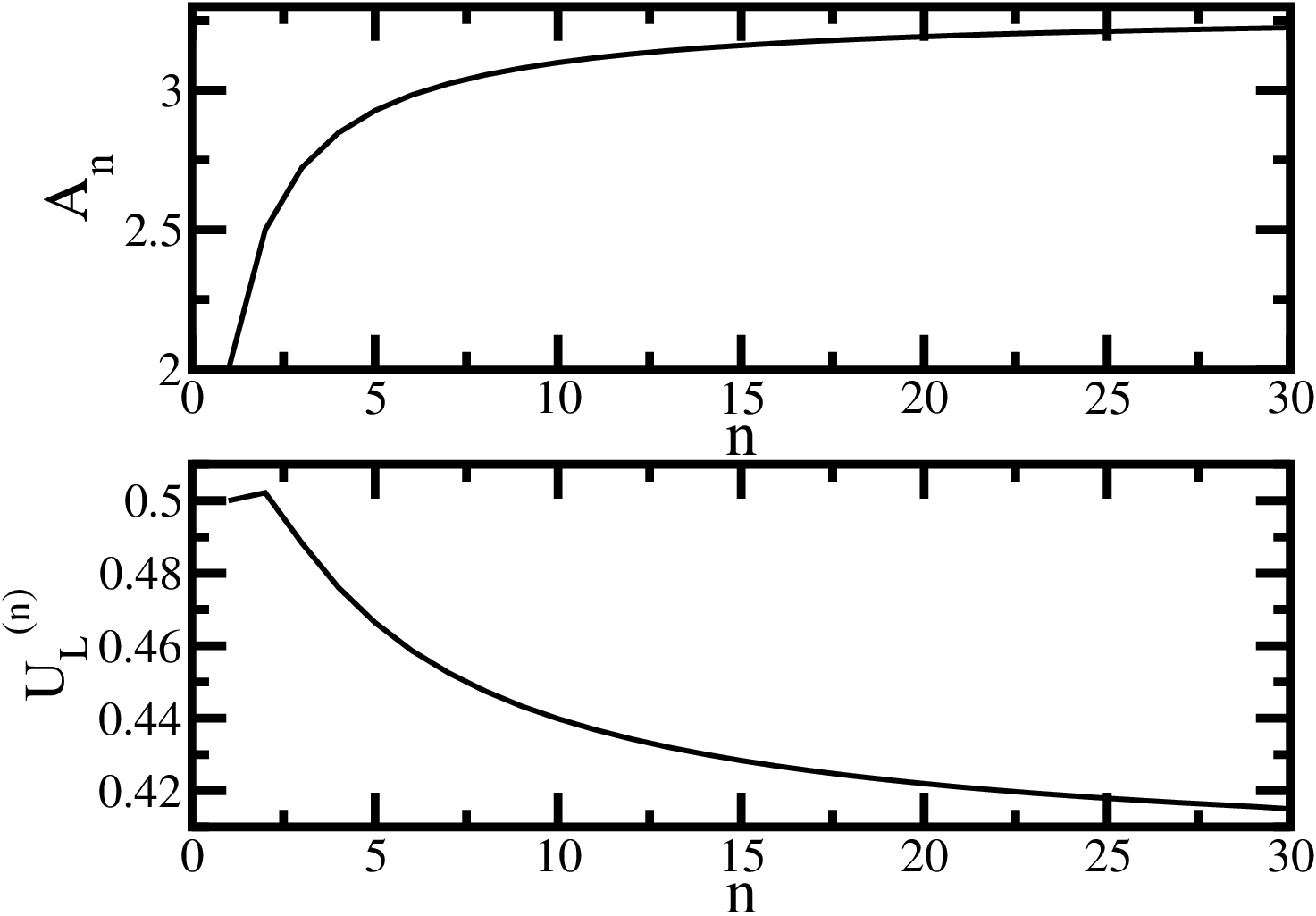}
 \caption{$A_n$ (defined in Eq. (\ref{eqn:C2_An})) as
     a function of $n$ (upper panel).  The Binder cumulant as a
     function of $n$ for a gapless system.}
 \label{fig:A_vs_n}
\end{figure}

The electronic contribution to the polarization can be obtained from
$Z_q$ by taking the first finite difference derivative with respect to
$q$.  Indeed, the electronic contribution to the many-body
polarization expression derived by Resta~\cite{Resta98} for
interacting systems can be derived by taking the first finite
difference derivative of $\ln Z_q$ at $q=0$,
\begin{equation}
   \langle X \rangle = i \frac{L}{2 \pi} \left(\ln Z_1 - \ln Z_0
   \right) = \frac{L}{2 \pi} \mbox{Im} \ln Z_1.
\end{equation}
To obtain the full polarization, the nuclear
contribution~\cite{Souza00,Resta00} also has to be added.  In most
models of interest (tight-binding type models) the nuclei are taken as
fixed in position.  Therefore, the nuclei do make a contribution to
the first cumulant (the polarization), but since higher order
cumulants describe fluctuations, the nuclei make no contribution.

We now show that for a band insulator, it is possible to relate $Z_q$,
as defined in Eq. (\ref{eqn:Z_q}) to the generalized Bargmann
invariant of Eq. (\ref{eqn:Z_qa}).  We can follow the steps of
Resta~\cite{Resta98}, and introduce a system periodic in $L$, with
lattice constant taken to be unity.  We will consider a system with
one filled band, but we will comment on the many-band generalization
below.  The $L$ Bloch vectors in the reciprocal cell $[0,2\pi/L)$ can
  be labelled,
\begin{equation}
  k_s = \frac{2 \pi}{L} s,s = 0,...,L-1,
\end{equation}
and the one-body orbitals will have the form
\begin{equation}
  \psi_{k_s}(x) = e^{i k_s x} u_{k_s}(x),
\end{equation}
where $u_{k_s}(x)$ is a periodic Bloch function.  The entire wave
function can be written as
\begin{equation}
  |\Psi_0 \rangle = \mathbb{A} \prod_{s=0}^{L-1} \psi_{k_s},
\end{equation}
where $\mathbb{A}$ is the anti-symmetrization operator.  Using this
wave function to write $Z_q$ (Eq. (\ref{eqn:Z_q})) results in
\begin{equation}
  \label{eqn:Z_qx}
  Z_q^{(b)} = \prod_{s=1}^L S(k_s,k_{s+q}),
\end{equation}
where $Z_q^{(b)}$ indicates the quantity in Eq. (\ref{eqn:Z_q})
averaged over a single Slater determinant representing an electronic
band, and where
\begin{equation}
  S(k_s,k_{s+q}) = \langle u_{k_s} | u_{k_{s+q}} \rangle,
\end{equation}
hence, it takes the form of Eq. (\ref{eqn:Z_qa}).

The extension to more than one filled band turns $S(k_s,k_{s+q})$ into
a matrix, and $Z_q$ into a determinant of the product of matrices,
\begin{equation}
  Z_q^{(b)} = \mbox{Det}\prod_{s=1}^L S(k_s,k_{s+q}),
\end{equation}
where
\begin{equation}
  S(k_s,k_{s+q}) = \langle u_{k_s,m} | u_{k_{s+q},m'} \rangle,
\end{equation}
where $m$ and $m'$ are band indices.  For $q=1$, we recover the known
quantity used to
derive~\cite{King-Smith93,Resta94,Resta00,Vanderbilt18} the MTP.  For
reference, and as a partial summary, in table I we summarize the
different types of characteristic functions encountered in this work.

We now analyze the implementation of he finite difference derivatives
in the MTP.  We first quote the Resta-Sorella expression for the
variance,
\begin{equation}
  \label{eqn:C2_RS1}
  C_2^{(RS1)} = - \frac{L^2}{2 \pi^2 } \mbox{Re} \ln Z_1,
\end{equation}  
according to the lowest order finite difference approximation (RS1).
This scheme has error $\mathcal{O}\left( L^{-2} \right)$.  An error of
order $\mathcal{O}\left( L^{-4} \right)$ can be achieved by using a
higher order finite difference derivative,
\begin{equation}
  \label{eqn:C2_RS2}
  C_2^{(RS2)} = \frac{L^2}{24 \pi^2 } \left[ \mbox{Re} \ln Z_2 - 16
    \mbox{Re} \ln Z_1\right].
\end{equation}

We would like to point out that when the expression $C_2^{(RS1)}$ (or
higher order finite difference approximations, such as $C_2^{(RS2)}$)
is applied to analyze quantum phase transitions, which are accompanied
by gap closure, a problem is encountered.  In the case of an ideal
conductor the twist operator shifts all the momenta of the system
producing a state which is orthogonal to the ground state:
\begin{equation}
  |\Psi_0\rangle \perp \exp \left( i \frac{2 \pi q}{L} \hat{X}
  \right)|\Psi_0\rangle,
\end{equation}
for $q \neq 0$.  As a result, at gap closure points $Z_q = 0$ for $q
\neq 0$.  As such, the variance, as approximated in
Eqs. (\ref{eqn:C2_RS1}) or (\ref{eqn:C2_RS2}), which depends on $\ln
Z_q$, will diverge in such a way that scaling information with respect
to system size $L$ is lost.

Another approximation scheme~\cite{Hetenyi19,Hetenyi20}, can also be
used, where we first remove the phases of $Z_q$, and then express
$M_2(=C_2)$ as,
\begin{equation}
  \label{eqn:C2_1}
  \tilde{M}_2^{(1)} = \tilde{C}_2^{(1)} = \frac{L^2}{2 \pi^2}\left(1 - |Z_1|\right).
\end{equation}
Since the mean polarization corresponds to the phase of $Z_1$, the
fact that the phases of $Z_q$ have been removed makes it explicit that
the statistical cumulants used to construct our Binder cumulants are
independent of the mean polarization or the origin of the coordinate
system (in other words, we are using central moments to construct
cumulants).

We introduced the
notation $\tilde{C}_2^{(1)}$ for the variance according to this
approximation scheme.  The superscript in parentheses refers to the
fact that the second derivative used is the lowest order finite
difference approximation.  The error in this scheme is
$\mathcal{O}\left( L^{-2} \right)$.  If the next order approximation
is used, the expression for the variance is
\begin{equation}
  \label{eqn:C2_2}
  \tilde{M}_2^{(2)} = \tilde{C}_2^{(2)} = \frac{L^2}{24 \pi^2}\left(|Z_2| -
  16|Z_1| + 15\right).
\end{equation}
The error in $\tilde{C}_2^{(2)}$ is $\mathcal{O}\left( L^{-4} \right)$.

A crucial difference between the variance expressions $C_2^{(RS1)}$,
$C_2^{(RS2)}$ and $\tilde{C}_2^{(1)}$, $\tilde{C}_2^{(2)}$ is that in
the latter set, the finite size scaling exponent of an ideal conductor
can be expected~\cite{Chiappe18} to be two since $Z_q \rightarrow 0$
(for $q \neq 0$), whereas in the former, the variance diverges even
for a finite system, which is unphysical.  Two is also the true upper
bound~\cite{Kudinov91,Chiappe18} for the finite size scaling exponent
for a closed gap system.  In contrast, the exponent for $C_2^{(RS1)}$,
$C_2^{(RS2)}$ is not bounded above.
\begin{figure}[ht]
 \centering
 \includegraphics[width=6cm,keepaspectratio=true]{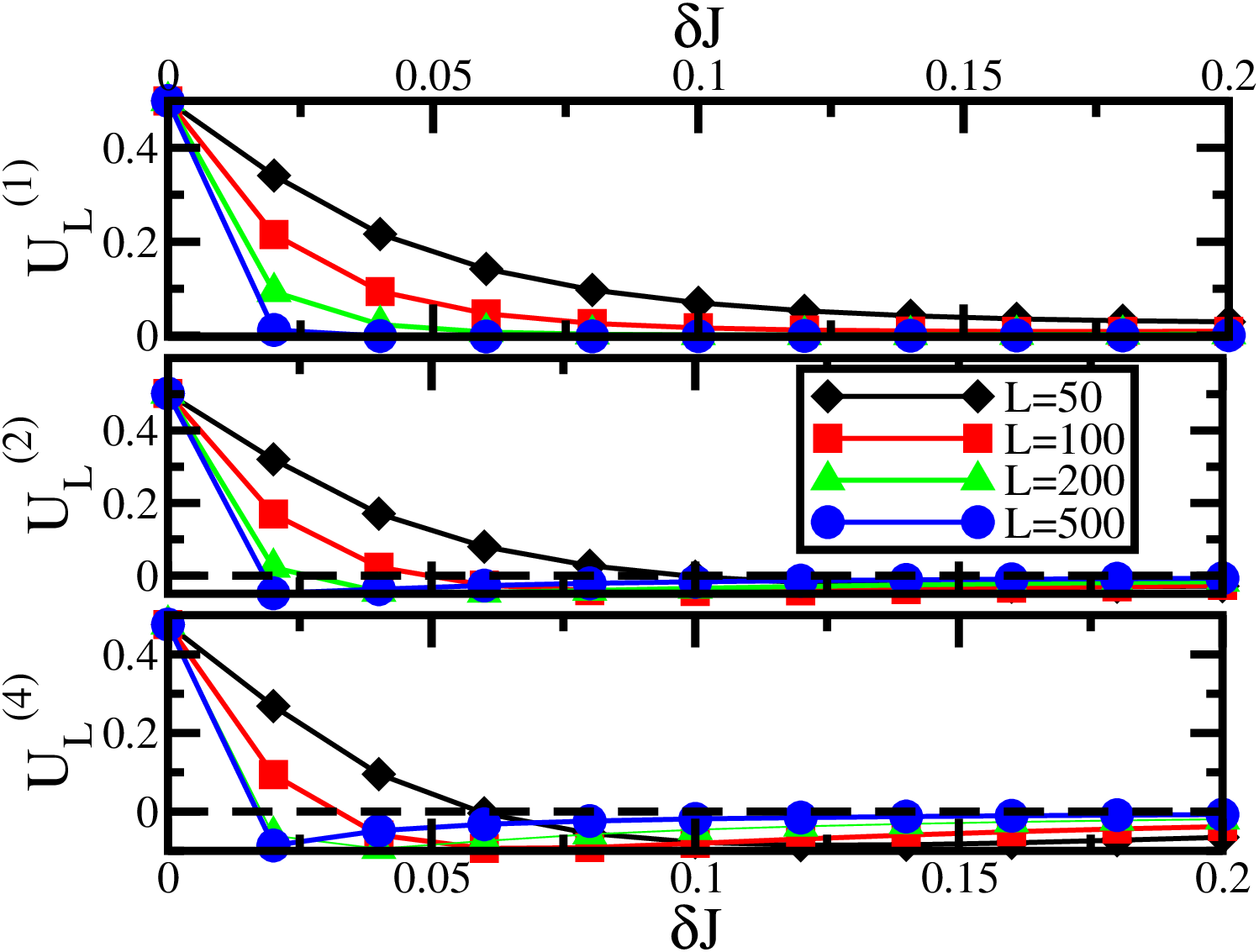}
 \caption{Binder cumulants $U_L^{(n)}$ approximated to different
   orders ($\mathcal{O}(L^{-2n}$)) for the one-dimensional SSH model
   (Eq. (\ref{eqn:SSH})).  The phase transition point occurs at $\delta J
   = 0$.  The different curves in each panel are for different system
   sizes $L=50,100,200,500$. }
 \label{fig:U_L}
\end{figure}

To construct a GBC, first the fourth moment of the
polarization is needed, which according to Eq. (\ref{eqn:M_n}) is the
fourth finite difference derivative of the characteristic function,
\begin{equation}
  \label{eqn:M4_1}
  \tilde{M}_4^{(1)} = \frac{L^4}{(2 \pi)^4} (2 |Z_2| - 8 |Z_1| + 6).
\end{equation}
The Binder cumulant can be written as
\begin{equation}
  \label{eqn:UL1}
  U_L^{(1)} = 1 -  \frac{\tilde{M}_4^{(1)}}{3[\tilde{M}_2^{(1)}]^2}.
\end{equation}
This quantity is essentially the same as the $U_4$ defined in
Eq. (\ref{eqn:U4}), albeit here it is introduced in the context of
MTP.  In general, the quantity $U_L^{(m)}$ for any $m$ is size
independent by construction, since $Z_q = 0$ for $q \neq 0$.  In
Eq. (\ref{eqn:UL1}) the subscript represents the system size, $L$, the
superscript represents the order of approximation used in calculating
the moments on the right hand side of the equation.  $U_L^{(1)}$ is
known as a fourth order Binder cumulant (because of the presence of
$\tilde{M}_4^{(1)}$).  Since in Eq. (\ref{eqn:UL1}) it is the moments
which are expressed in terms of a finite difference derivative, we
will refer to this way of calculating $U_L^{(1)}$ as the moment based
approximation (MBA).

Below we make comparisons between the GBC ($U_L^{(1)}$) calculated
based on the MBA and based on the Resta-Sorella approach, which
amounts to taking the finite difference derivative of $\ln Z_q$.  For
the fourth cumulant, the Resta-Sorella approach (up to
$\mathcal{O}\left( L^{-2} \right)$) gives
\begin{equation}
  \label{eqn:C4_RS1}
  C_4^{(RS1)} = \frac{L^4}{8 \pi^4 } \left[ \mbox{Re} \ln Z_2 - 4
    \mbox{Re} \ln Z_1\right].
\end{equation}
Using this, we can write the $\mathcal{O}\left( L^{-2} \right)$
approximation to the fourth order GBC as
\begin{equation}
  \label{eqn:UL1_RS1}
    U_L^{(RS1)} = -  \frac{C_4^{(RS1)}}{3 [C_2^{(RS1)}]^2}.
\end{equation}
Binder cumulants correct up to higher orders can be constructed using
higher order finite difference derivatives.
\begin{figure}[ht]
 \centering
 \includegraphics[width=6cm,keepaspectratio=true]{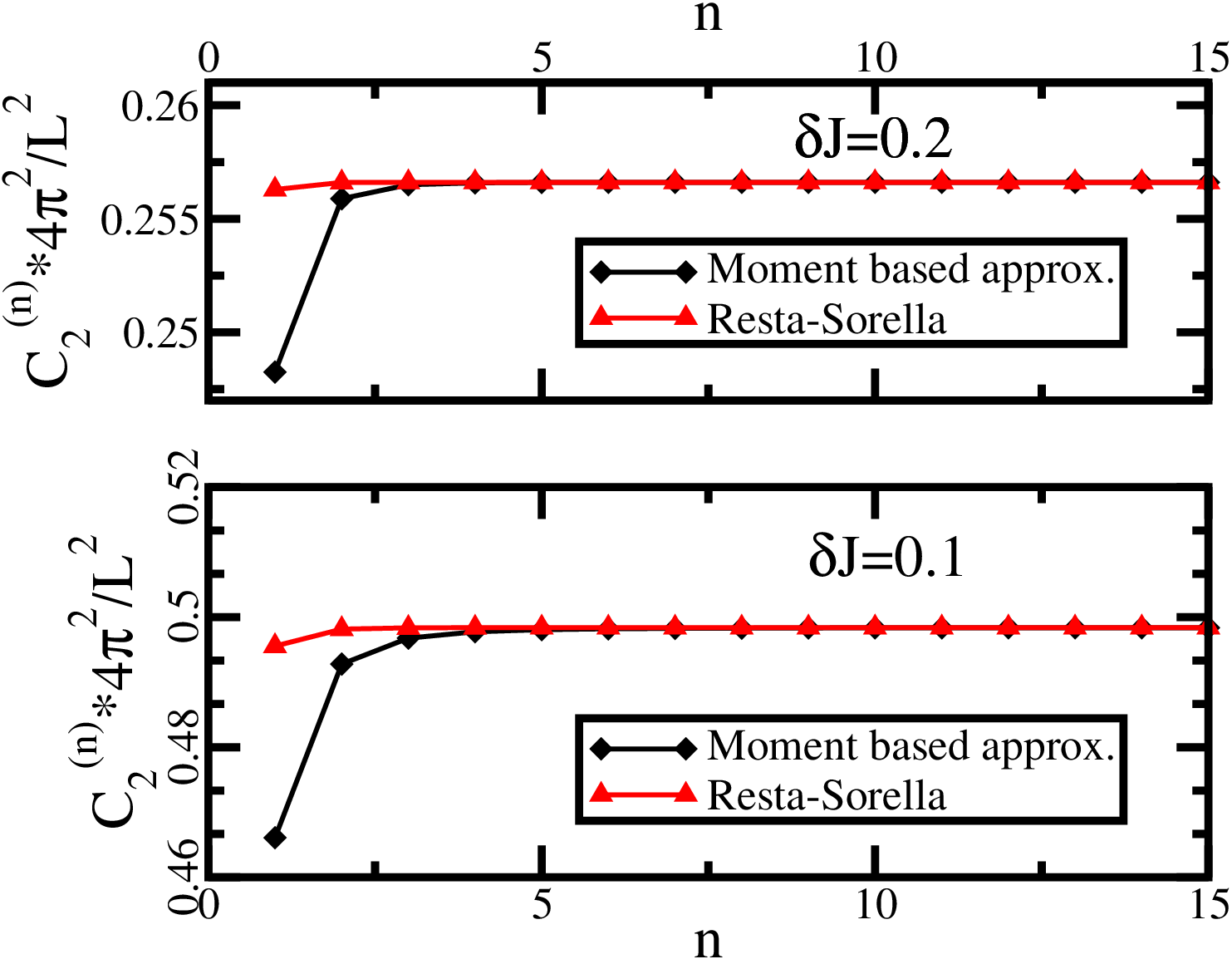}
 \caption{Variance as a function of the order of approximation, $n$,
   one-dimensional SSH model (Eq. (\ref{eqn:SSH})), two cases, $\delta J =
   0.2$, and $\delta J = 0.1$ calculated via two methods: the
   Resta-Sorella approximation, and the moment based approximation
   (MBA) developed in this work.}
 \label{fig:X2_n}
\end{figure}

\section{Approximating gap closure}
\label{sec:ideal_conductor}

In this section we investigate how the MBA expressions for the
variance and the GBC behave in the absence of an energy gap.  We
recall that when the gap is open the GBC is zero (in the thermodynamic
limit, which is the same as $\delta t \rightarrow 0$ in
Eq. (\ref{eqn:U4Gamma})), however, finite values result in closed gap
regions.  Since our expressions are finite difference approximations,
the question that arises is what can we expect at approximations of
higher order?

We have shown above that at gap closure $Z_q = 0$, except for $Z_0 =
1$.  We can obtain higher order approximations of the variance by
taking higher order finite difference approximations to the second
derivative.  The variance for the first two orders of approximation
are given in Eqs. (\ref{eqn:C2_1}) and (\ref{eqn:C2_2})
($\mathcal{O}\left( L^{-2} \right)$ and $\mathcal{O}\left( L^{-4}
\right)$, respectively).  In general, the variance will have the form
\begin{equation}
  \label{eqn:C2_An}
  C_2^{(n)} = A_n \frac{L^2}{4 \pi^2},
\end{equation}
where $A_n$ depends on the order of the approximation.  For $n=2$,
$A_n = \frac{5}{2}$ (compare with Eq. (\ref{eqn:C2_2})).  For $n$ up
to 30, the value of $A_n$ is shown in Fig. \ref{fig:A_vs_n}.  The
parameter $A_n$ appears to converge as a function of $n$.  We can
estimate a lower bound of $A_{30}=3.224$ for $A_\infty$.

The crucial point is that while the value of the variance varies with
level of approximation, the scaling with system size within MBA is
always two.  This also holds for higher order cumulants.  For the
GBCs, this means that while the value of the GBC at gap closure points
will vary as a function of the order of approximation, a GBC of a
given approximation will still be size independent at gap closure
points.  In this sense, the GBCs at a given level of approximation are
universal.

The lower panel of Fig. \ref{fig:A_vs_n} shows the GBC as a function
of the order of the approximation, $n$.  The value of the cumulant
varies considerably as a function of approximation, it decreases as
$n$ is increased.  The curve appears to level off to a constant value.
Locating quantum phase transition points is possible at any order of
approximation, since size independence of the GBC is still guaranteed.

\section{Model calculations}
\label{sec:Model_Calculations}
\subsection{One-dimensional Su-Schrieffer-Heeger model}
\label{ssec:1DSSH}

The Hamiltonian of the Su-Schrieffer-Heeger model~\cite{Su79} is given
by,
\begin{equation}
  \label{eqn:SSH}
  \hat{H} = \sum_{i=1}^L (-J c_i^\dagger d_i - J' d_i^\dagger c_{i+1} + \mbox{H.c.}),
\end{equation}
where $c_i^\dagger$($d_i^\dagger$) denote creation operators in one
unit cell $i$ on sublattice $A$($B$).  This model already has a long
history, here we emphasize that it is insulating for $J'\neq J$, and
gap closure occurs at $J'=J$.  For convenience we will use the
parametrization $J = \bar{J} + \delta J$, $J' = \bar{J} - \delta J$,
hence, the gap closure occurs at $\delta J = 0$.  We will take
$\bar{J} = 1$.

Fig. \ref{fig:U_L} shows the Binder cumulant based on approximations
of different order ($n=1,2,4$) for different system sizes for the 1D
SSH model near the gap closure point.  Indeed, the Binder cumulants
are independent of system size at the phase transition point, however,
they exhibit considerable size dependence away from it.

In Fig. \ref{fig:X2_n} the normalized variance is shown as a function
of the order of approximation for the two cases ($\delta J = 0.1$ and
$\delta J = 0.2$) for both the MBA, and the Resta-Sorella type
approximation carried to higher orders.  While, for the gapless case
(Fig. \ref{fig:A_vs_n}) the approximation is crucial to obtain
quantitative results, for the gapped system the convergence is rapid,
a calculation correct up to order $\mathcal{O}\left( L^{-8} \right)$
($n=4$) is already converged.  It is to be noted, however, that the
usual Resta-Sorella calculation does give faster convergence as a
function of approximation.  Still, due to the logarithm, at gap
closure points it is undefined, so size scaling information is lost.

While the variance at order $\mathcal{O}\left( L^{-2} \right)$ is not
necessarily converged, phase transition points will still be detected,
because the GBCs are size independent at each level of finite
difference approximation.  At $n=1$ ($\mathcal{O}\left( L^{-2}
\right)$), the universal value of the GBC at gap closure is $\frac{1}{2}$, as
$n$ is increased, the values will be differ, according to
Fig. \ref{fig:A_vs_n} (lower panel).  Increasing the order of the
finite difference approximation is only important if accurate
variance, skew, kurtosis, etc. is desired.
\begin{figure}[ht]
 \centering
 \includegraphics[width=6cm,keepaspectratio=true]{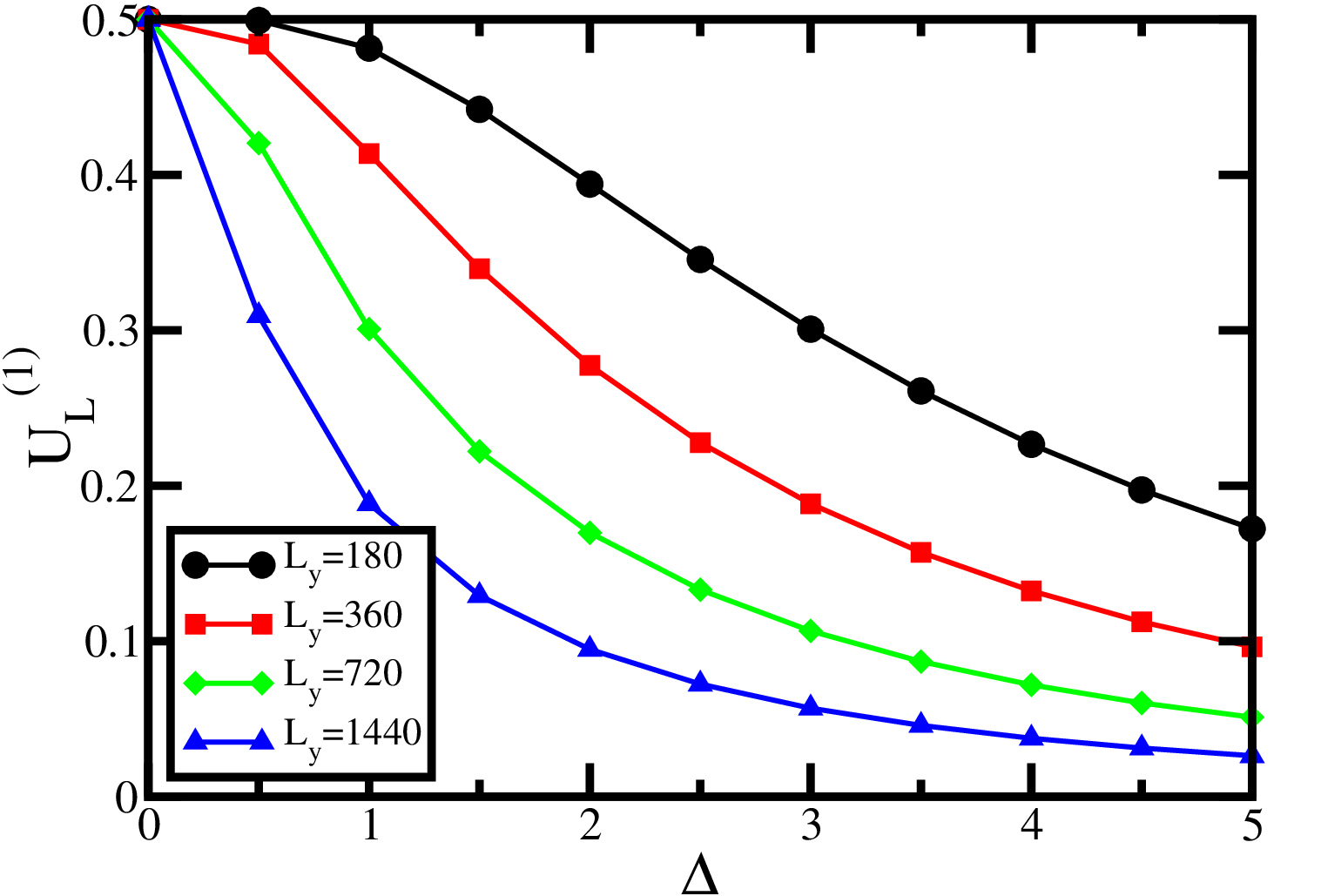}
 \includegraphics[width=6cm,keepaspectratio=true]{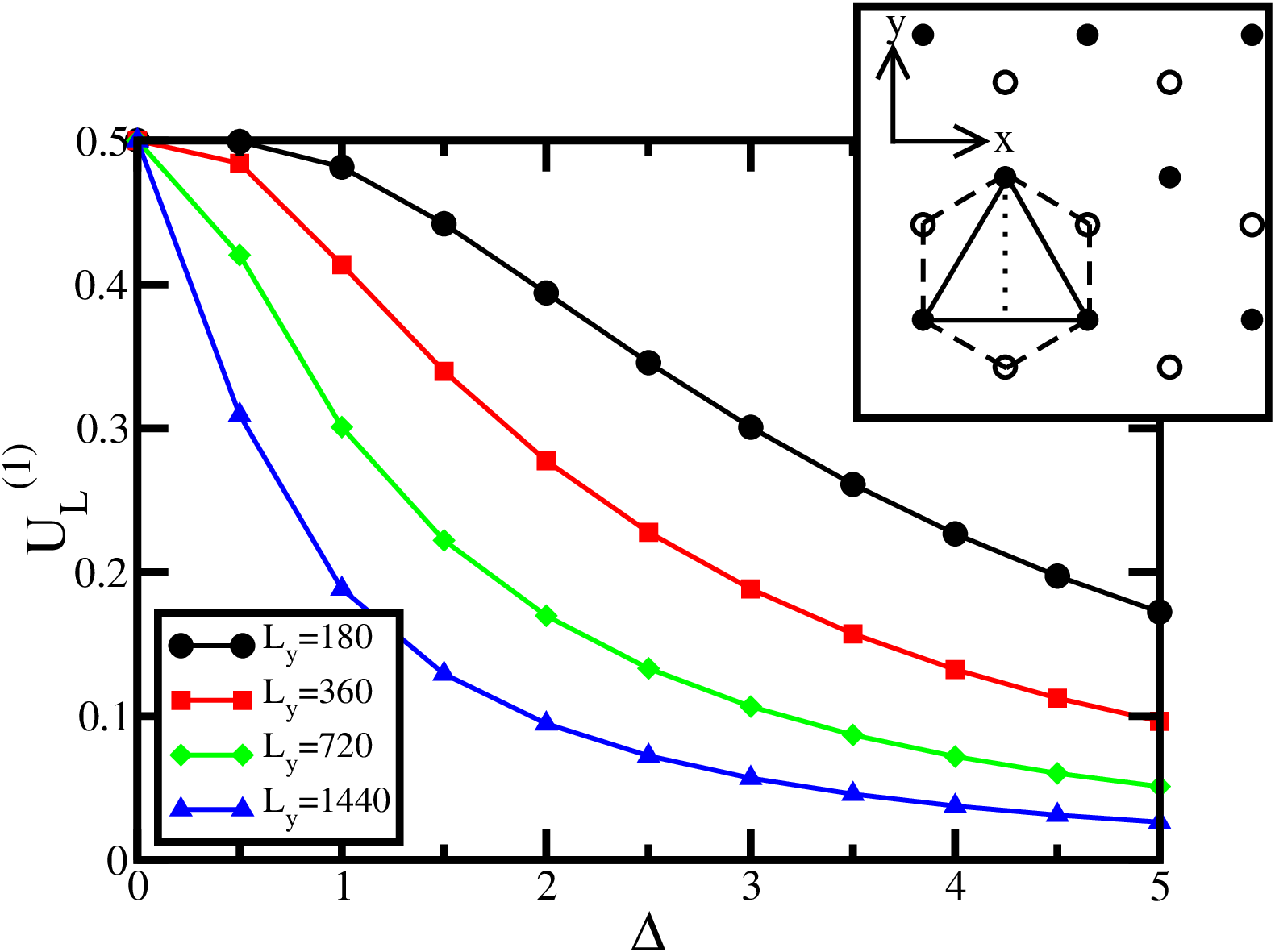}
 \includegraphics[width=6cm,keepaspectratio=true]{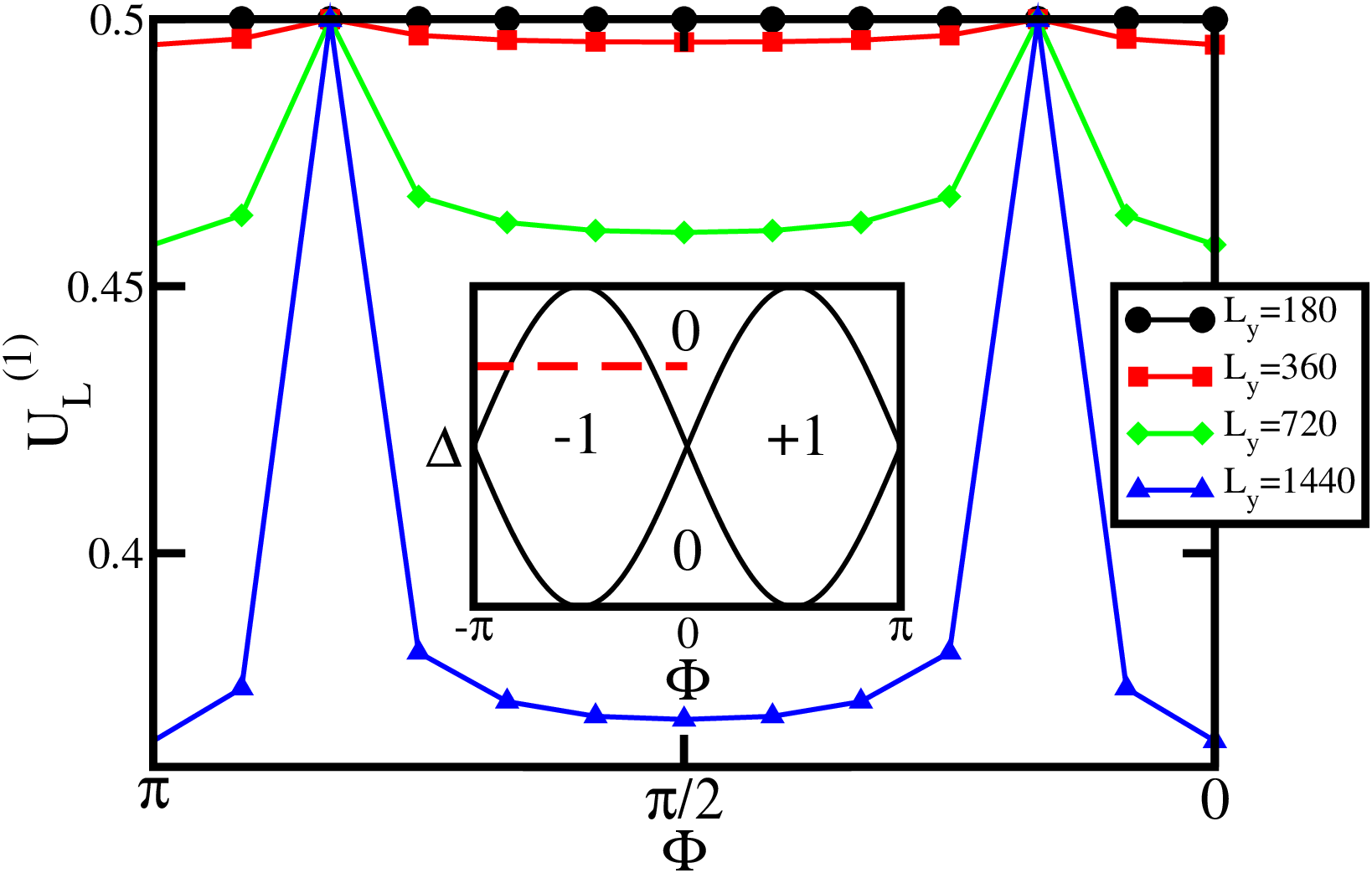}
 \caption{Geometric Binder cumulants for two-dimensional models.
   Upper panel shows the results for a tight-binding model with an
   on-site potential (strength $\Delta$) on a square lattice, the
   middle panel shows the results for a tight-binding model with an
   on-site potential (strength $\Delta$) on a hexagonal lattice, the
   lower panel shows results for the topological Haldane model on a
   hexagonal lattice.}
 \label{fig:U_L_2D}
\end{figure}

\subsection{Two dimensional models}
\label{sec:2D}

In two dimensions we consider two different cases of gap closure.  For
the most basic case (square lattice, tight-binding) gap closure occurs
at points along a closed curve within the Brillouin zone.  Gap opening
can be achieved, for example, by an alternating on-site potential.  If
the same system is placed on a honeycomb lattice, the gap closure
occurs only at isolated points in the Brillouin zone (Dirac
points).

For the square lattice, we take the Hamiltonian to consist of nearest
neighbor hopping and an on-site potential of strength $\Delta$,
\begin{equation}
  \hat{V} = \Delta \sum_{j,k} (-1)^{j+k} \hat{n}_{j,k}.
\end{equation}
In reciprocal space the model can be written as:
\begin{equation}
  \label{eqn:Hkxky}
  \hat{H} = \sum_{k_x,k_y} \left[\epsilon(k_x,k_y) \sigma_x + \Delta \sigma_z \right],
\end{equation}
where 
\begin{equation}
  \epsilon(k_x,k_y) = - 2 \cos(k_x) - 2 \cos(k_y),
\end{equation}
and where $\sigma_x$ and $\sigma_z$ denote Pauli matrices.  The
tight-binding hopping parameter was taken to be unity.  We take the
basis vectors of the lattice (which give the positions of the $A$
sublattice) to be,
\begin{equation}
  \vec{\bf a}_1 = (2,0);\vec{\bf a}_2 = (1,1),
\end{equation}
leading to a two-dimensional $k$-space grid of
\begin{equation}
  k_x = \frac{\pi}{L_x} m_x,  k_y = \frac{2 \pi}{L_y} m_y - \frac{\pi}{L_x} m_x, 
\end{equation}
with $m_x,m_y = 0,...,L-1$.  We define the two-dimensional
polarization amplitude as
\begin{equation}
  \label{eqn:Z_q_2D}
  Z_q = \prod_{m_x,m_y} \langle m_x,m_y | m_x, m_y + q \rangle,
\end{equation}
where $|m_x,m_y\rangle$ is an eigenstate of $\epsilon(k_x,k_y) {\bf
  \sigma_x} + \Delta {\bf \sigma_z}$.  Using $Z_q$ we can calculate
GBCs as given in Eqs. (\ref{eqn:C2_1}), (\ref{eqn:M4_1}),
and (\ref{eqn:UL1}).

The upper panel of Fig. \ref{fig:U_L_2D} shows the GBC for different
system sizes.  The lattice size was held fixed in the $x$-direction
($L_x = 180$), and scaling was done only in the $y$-direction, as
indicated by the different $L_y$ values in the figure.  At the gap
closure point, all GBCs converge to their calculated value of $\frac{1}{2}$,
away from this point they exhibit size dependence, and decrease
monotonically.
\begin{figure}[ht]
 \centering
 \includegraphics[width=6cm,keepaspectratio=true]{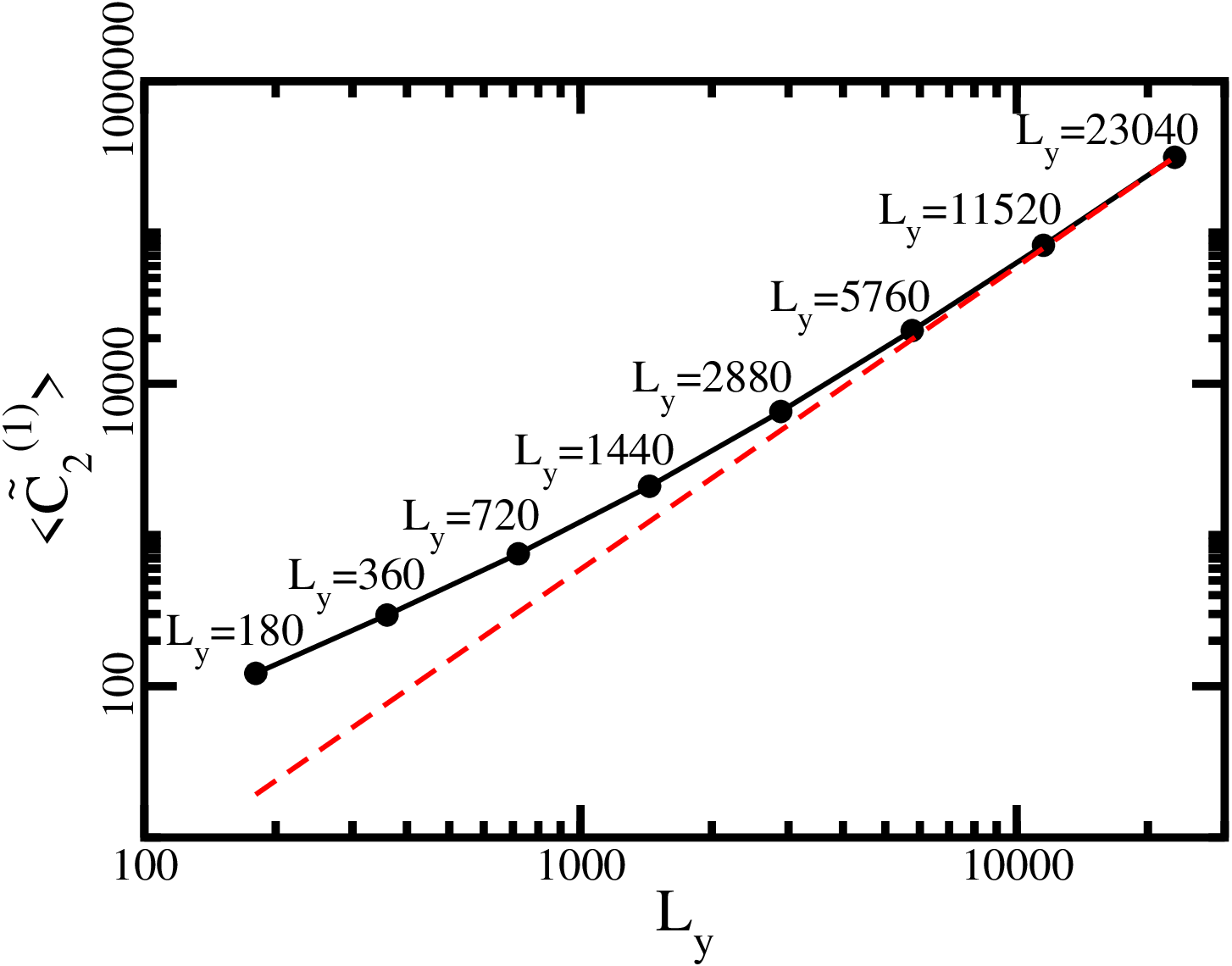}
 \includegraphics[width=6cm,keepaspectratio=true]{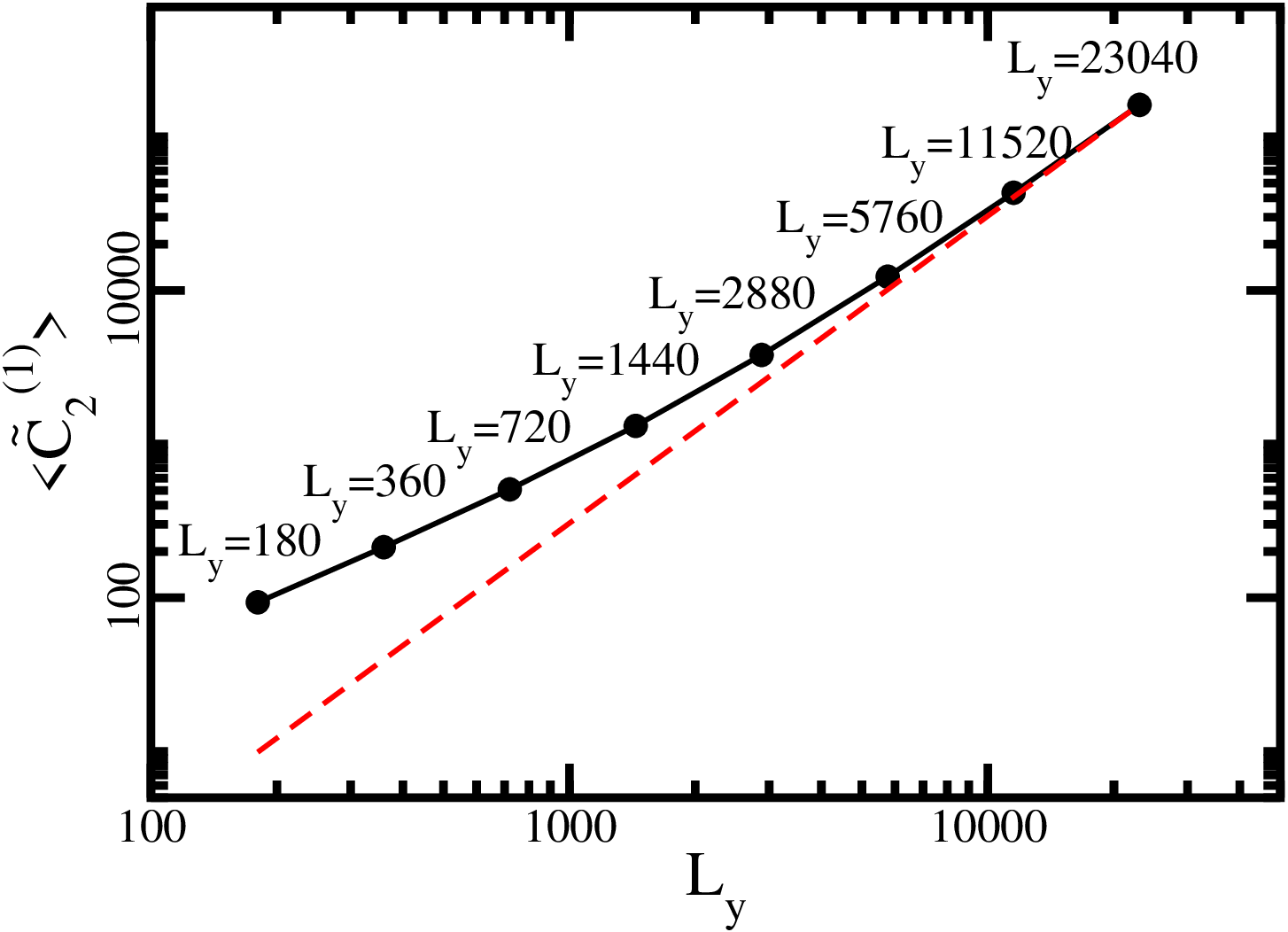}
 \caption{Variance for the hexagonal lattice (upper panel), and the
   Haldane model (lower panel), both at gap closure, averaged in the
   transverse direction, shown on a log-log plot.  For all
   calculations $L_x$ is held fixed, and $L_y$ is varied.  The values
   of $L_y$ are shown above each data point.  The straight dashed line
   is the log-log plot of a curve of the form $f(L_y) = a L_y^2$.  For
   the hexagonal model (upper panel) $a=0.000593473$, for the Haldane
   model (lower panel) $a=0.000325063$.}
 \label{fig:C2loglog}
\end{figure}

We apply exactly the above procedure for the hexagonal lattice with an
alternating on-site potential.  In this case, the Hamiltonian takes
the form,
\begin{equation}
  H =   \left[ {\begin{array}{cc}
    \Delta & f \\
    f^* & -\Delta \\
  \end{array} } \right].
\end{equation}
where
\begin{equation}
  f = \exp\left(i \frac{k_y}{\sqrt{3}}\right) + 2 \cos
  \left( \frac{k_x}{2}\right) \exp\left(-i
  \frac{k_y}{2\sqrt{3}}\right).
\end{equation}
The coordinate system we use is shown in the inset of the middle panel
of Fig. \ref{fig:U_L_2D}.  The equilateral triangle connecting second
nearest neighbors has sides of length unity.  The main figure of the
middle panel of Fig. \ref{fig:U_L_2D} shows the GBCs for the hexagonal
lattice.  Again, in the gapped region ($\Delta>0$) the GBCs are size
dependent, at the gap closure point, ($\Delta=0$) all the GBCs
converge to a value of $\frac{1}{2}$.

For completeness we also investigated the topological Haldane
model~\cite{Haldane88,Thonhauser06}.  In this model, in addition to
the on-site potential (which breaks inversion symmetry), a
time-reversal symmetry breaking term is also added, consisting of a
magnetic flux which encircles the center of each hexagon in such a
ways that the Peierls phases ($\Phi$) of the flux only affect the
second nearest neighbor hoppings.  The tuning of the on-site potential
and phase on second nearest neighbor hoppings renders possible the
independent control (opening or closing) of each Dirac point
separately.  The Hamiltonian now has the form,
\begin{equation}
  H =   \left[ {\begin{array}{cc}
    g & f \\
    f^* & -g \\
  \end{array} } \right],
\end{equation}
with
\begin{equation}
  g = \Delta + 2t_2 \sin(\Phi)\left[ \sin(k_x) + 2 \sin\left(\frac{k_x}{2}\right)\cos\left(\frac{\sqrt{3}k_x}{2}\right)\right].
\end{equation}
The lower panel of Fig. \ref{fig:U_L_2D} shows our results for the
Haldane model.  The inset of the lower panel shows the phase diagram
of the model (solid line), and the dashed line indicates the line
along which the GBCs were calculated.  The main figure of the lower
panel shows the GBCs themselves.  Two phase transition points are
encountered, where the GBCs all take a value of $\frac{1}{2}$. In
between and outside the phase transition points, the GBCs are strongly
size dependent.

The Fermi surface of the square lattice and the honeycomb lattice are
different, the former is a closed one-dimensional curve (a square in
the Brillouin zone), the latter is zero-dimensional (two Dirac
points).  An interesting question is whether finite size scaling
methods can distinguish between the two.  The quantity capable of
doing this is the variance of the polarization averaged in the
transverse direction.  We define:
\begin{equation}
  \label{eqn:C2_av}
  \langle \tilde{C}_2^{(1)} \rangle = \frac{1}{L_x} \sum_{k_x}  \tilde{C}_2(k_x),
\end{equation}
where $C_2(k_x)$ is the variance of the polarization in the
$y$-direction for a fixed value of $k_x$ (according to the definition
in Eq. (\ref{eqn:C2_1})).  Of particular interest is the scaling of
$\langle C_2^{(1)} \rangle$ as a function of system size.  We
approximate
\begin{equation}
\langle \tilde{C}_2^{(1)} \rangle = a L^\gamma,
\end{equation}
where $a$ is a coefficient, and $\gamma$ is the size scaling exponent.
For the square lattice, when the system size is scaled both in the $x$
and $y$ directions simultaneously, the size scaling exponent
$\gamma=2$ at the gap closure point, and $\gamma=1$ for gapped
systems.  We checked this via calculations for system sizes of
$L_x=L_y=180\times 2^k$, $k=0,1,2,3,4,5,6,7$.  For the hexagonal
lattice, based on calculations of the same set of system sizes, when
scaling is done in both directions $\gamma \approx 1.12$ for the
gapless case, while $\gamma=1$ for gapped systems.  However, if
scaling is only done in one dimension, then for large system sizes the
scaling exponent becomes $\gamma = 2$ for the gapless case (for the
gapped systems it still holds that $\gamma=1$).  The results are shown
in Fig. \ref{fig:C2loglog}.  For $L_x=180$ fixed, as $L_y$ is varied,
the log-log plot for the average variance as a function of system size
does not show a straight line, however, as $L_y$ becomes large, the
function does tend to a curve with $L_y$ dependence $L_y^2$, hence
$\gamma=2$.  The different panels of Fig. \ref{fig:C2loglog}
correspond to the different hexagonal models: upper panel(lower panel)
shows the honeycomb tight-binding(Haldane) model.

These results can be explained from the form of Eq. (\ref{eqn:C2_av}).
For a two-dimensional system whose Fermi surface is a closed curve
which runs over the entire Brillouin zone, a gap closure point will be
encountered for each $k_x$, which will render $Z_1$ zero, and lead to
a contribution to $\langle C_2^{(1)} \rangle$ with a scaling exponent
of $\gamma=2$.  For a system with Dirac points, the contributions to
$\langle C_2^{(1)} \rangle$ scale as $L_y$, except for values of
$k_x$, which encounter Dirac points, which scale as $L_y^2$.  In this
case $L_y$ has to be increased so that this one term dominates the
scaling of $\langle C_2^{(1)} \rangle$.
\begin{figure}[ht]
 \centering
 \includegraphics[width=6cm,keepaspectratio=true]{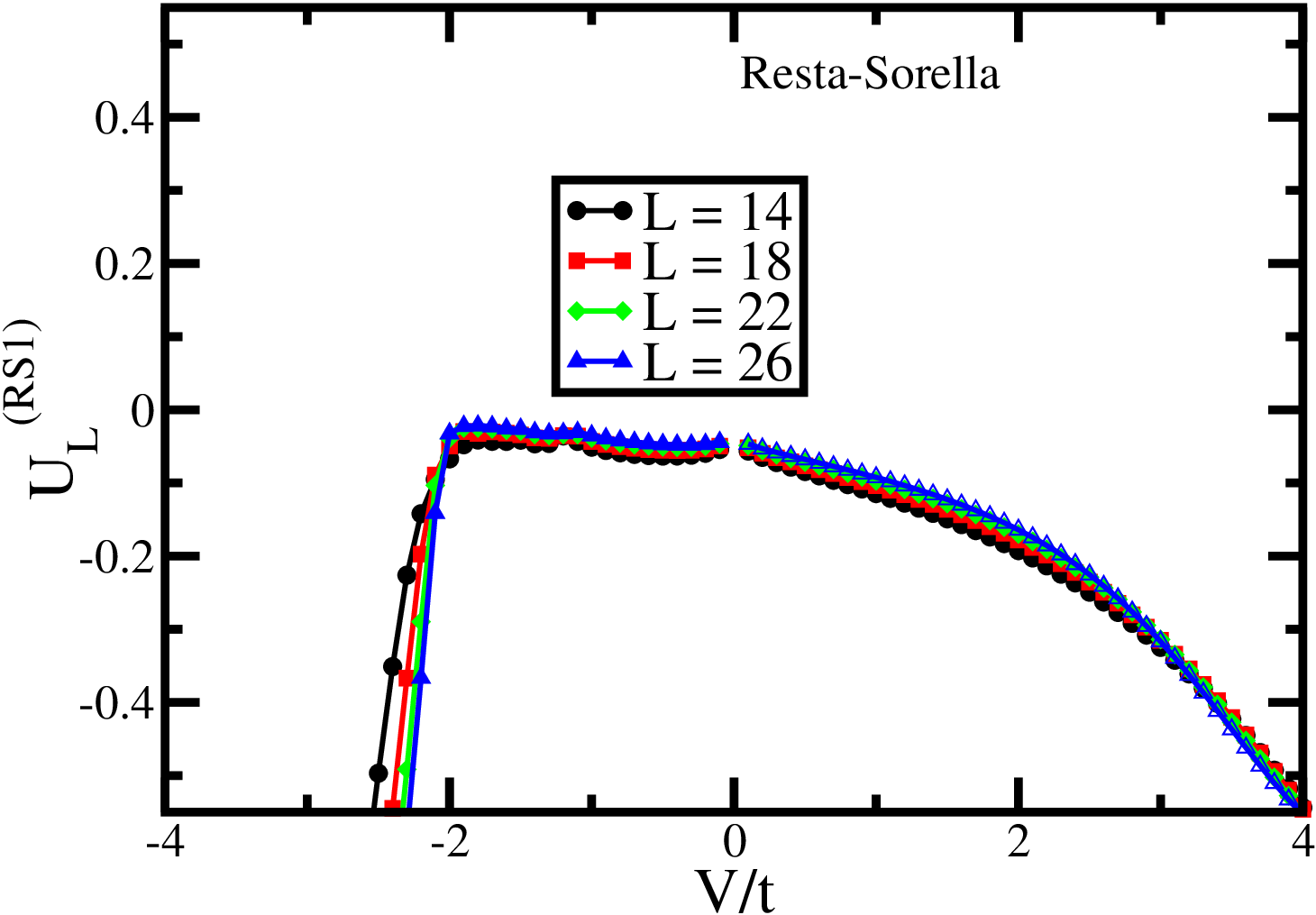}
 \includegraphics[width=6cm,keepaspectratio=true]{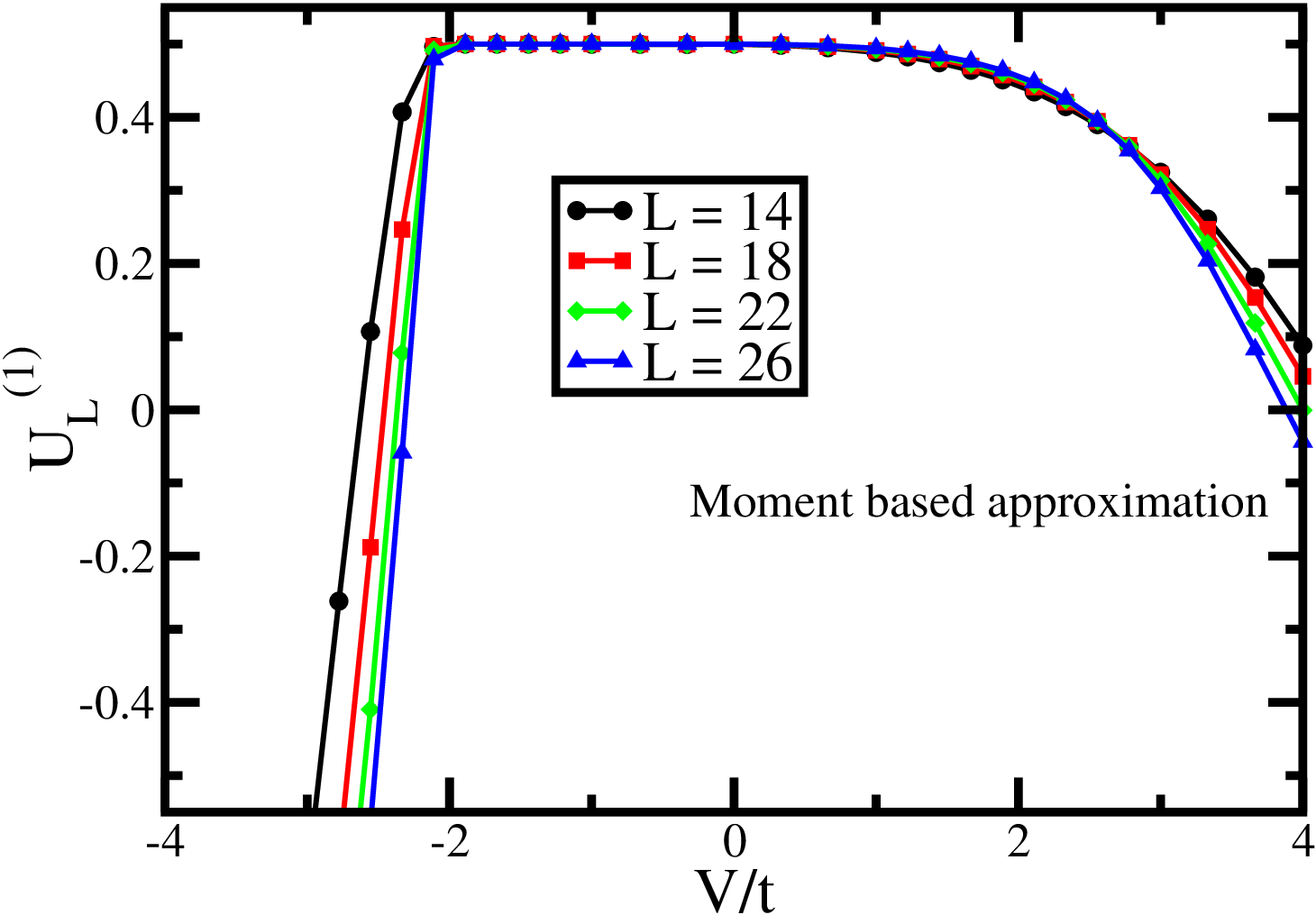}
 \caption{Binder cumulant for the $t-V$ one-dimesional interacting
   model, as a function of $V/t$.  The upper panel shows results for
   the Resta-Sorella approximation, the lower one for the moment based
   approximation.}
 \label{fig:bc_tV}
\end{figure}

\subsection{Correlated one-dimensional model}
\label{sec:1D_tV}

Our last model calculation is the $t-V$ model, a one-dimensional
tight-binding model (hopping $t$) with nearest neighbor interaction
$V$.  The Hamiltonian is given by
\begin{equation}
  H = \sum_{i=1}^L \left[ -t(c_i^\dagger c_{i+1} + c_{i+1}^\dagger
    c_i) + V n_i n_{i+1}\right].
\end{equation}
This model exhibits phase transitions at $V=\pm 2t$.  If $|V|<2t$ the
system is in a gapless Luttinger liquid phase, whereas, $|V|>2t$ is an
insulating phase.  The phase transition at $V=2t$ is second order,
while the one at $V=-2t$ is first order.  We perform exact
diagonalization calculations (Lánczos method).
\begin{table}[t]
\begin{center}
\begin{tabular}{ |c|c|c| } 
 \hline
 System sizes & Repulsive fixed point & Repulsive fixed point  \\ 
 \hline
 $L=16/8$ &  $V=3.00t$ & $V=-2.21t$ \\ 
 $L=20/10$ & $V=2.75t$ & $V=-2.13t$ \\ 
 $L=24/12$ & $V=2.62t$ & $V=-2.08t$ \\ 
 $L=28/14$ & $V=2.52t$ & $V=-2.06t$ \\
 \hline
\end{tabular}
\end{center}
    \caption{The repulsive fixed points for the $t-V$ model for different pairs of system sizes.}
    \label{tab:t-V}
\end{table}

In Fig. \ref{fig:bc_tV} the Binder cumulant is shown for a sweep
across both phase transitions for different system sizes.  The upper
panel shows results for the Resta-Sorella approximation (the $V=0$ is
missing, since there the logarithms give a divergent result), the
lower for MBA.  In both cases the first-order phase transition is easy
to identify.  Overall, it is difficult to say whether the
Resta-Sorella method converges to a precise value in the gapless
phase, since even there there is some size dependence, and the value
of the cumulant is difficult to identify.  In the MBA, the cumulants
converge in the region $-2t<V<2t$ to a value of one half, and the
phase transition at $V=-2t$ is clearly identifiable.  As for the
second order phase transition at $V=2t$, the MBA is superior to the
Resta-Sorella approach, partly because the Resta-Sorella method does
not converge to a universal value in the gapless phase.

Another striking feature of the results in Fig. \ref{fig:bc_tV} is
that in the ordered phases $V>|2t|$ the Binder cumulant is not zero.
This is due to the fact that in the many-body case the polarization is
not a usual Berry-Zak phase (integral across the Brillouin zone), it
is, instead, a single point~\cite{Resta00} Berry phase.  One can
calculate $Z_q$ in the ordered phases (see Ref. \cite{Hetenyi19}), and
one obtains a Binder cumulant of minus infinity.  Still, the gap
closure region is easily identified.

For comparison, we also used another method to locate the phase
transition point.  This method is an application of real-space
renormalization to the polarization amplitude~\cite{Hetenyi21}. For a
given value of $V$, $Z_1$ is calculated for some initial system size
$L$.  Then for $L/2$ the value of $V$ is tuned to reproduce the same
$Z_1$ as for the larger system.  This generates a renormalization
trajectory, via
\begin{equation}
  Z_1(V_{j+1},L/2) = Z_1(V_j,L),
\end{equation}
and fixed points as a function of $V$ can be determined.  The
trajectories tend to an attractive fixed points at $V=0$, $V=\infty$,
and $V=-\infty$, and two repulsive fixed points occur at finite $V$ in
between the three attractive fixed points.  The repulsive fixed points
correspond to the phase transition points.  Our calculated results for
the repulsive fixed points are shown in table \ref{tab:t-V}.  Again,
the phase transition point in the attractive region of the model
($V<0$) is easy to locate, our best result being $V = -2.06t$.  In the
region $V>0$, for the largest system size we are able to do, the
repulsive fixed point is at $V\approx2.52$, in good agreement with our
results in the lower panel of Fig. \ref{fig:bc_tV}.  We emphasize that
the limitation is not the finite size scaling methods advocated here,
but the system size accessible by exact diagonalization.

\section{Conclusion and Future Prospects}
\label{sec:cnclsn}

In this work we explored a generalization of the study of adiabatic
cycles initiated in the context of quantum mechanics by
Berry~\cite{Berry84}.  A geometric phase becomes non-trivial if the
adiabatic cycle over which it is defined encircles a degeneacy point.
If the path is modified so that it crosses the degeneracy point, the
Berry phase becomes undefined.  The direction of our generalization
was the consideration of cycles which cross isolated degeneracy
points.  We showed that it is possible to define quantities which are
physically well defined in this case.  The quantities which achieve
the above goal are ratios of the cumulants associated with adiabatic
cycles.  We gave a detailed analysis of how cumulant generating
functions can be defined in the general context of the geometric
phase, as well as the particular context of the modern theory of
polarization.  For the latter we drew on the work of Souza, Wilkens,
and Martin ~\cite{Souza00}.

In the modern polarization context we defined cumulant ratios which we
showed to be useful for finite size scaling of quantum phase
transitions.  We argued that the method often used in the literature,
due to Resta and Sorella~\cite{Resta98,Resta99}, has a drawback,
namely that the size scaling of the variance of the polarization is
lost at the phase transition points.  We presented an alternative
approximation scheme which leads to controlled divergence of the
system size at phase transition point.  The geometric Binder cumulants
we derived are ratios of cumulants of the polarization defined
according to our approximation scheme.

We also presented numerical model calculations of the techniques
developed.  The geometric Binder cumulant can detect the gap closure
in all cases studied, which included one-dimensional and
two-dimensional band models, and a one-dimensional correlated system.

In two dimensions we investigated two different types of gap closures.
Some models exhibit one-dimensional Fermi surfaces (for example, a
simple two-dimensional tight-binding model.  Other models, such as
graphene, or the topological Haldane model, exhibit isolated points as
Fermi surfaces (Dirac points).  We showed that the geometric Binder
cumulant detects gap closure in both cases, provided, that scaling is
done in one dimension only.  We also showed that the variance, if
system size scaling is done in two directions, will not detect the gap
closure in the case of models with Dirac points, because, even though
for one particular $k$-vector the gap closure does indeed cause a
divergence, this effect is statistically suppressed by the presence
other $k$-vectors.  Scaling in one direction solves this problem.

The geometric phase has by now a considerable history, and many
generalizations have been done.  One well-known
extension~\cite{Wilczek84} is the application to systems in which
degeneracies originating from symmetry exist, for example, if
time-reversal symmetry leads to Kramers doublets~\cite{AldenMead87}.
In this case the integral defining the geometric phase becomes the
Wilson loop, and the geometric phase itself becomes an $SU(2)$ matrix
(a quaternion).  The adiabatic loop is of particular interest if it
encircles crossing points between Kramers doublets.  We are confident
that our ideas are applicable to these types of more complex
scenarios.

\section*{Acknowledgments}
This research was supported by the National Research, Development and
Innovation Office (NKFIH), within the Quantum Technology National
Excellence Program (Project No. 2017-1.2.1-NKP-2017-00001) and
K142179.

\end{document}